\documentclass[twocolumn]{bmcart}

\usepackage[utf8]{inputenc} 
\usepackage{algorithm}
\usepackage[noend]{algpseudocode}
\usepackage{tabularx}
\usepackage{multirow}

\usepackage[pdftex]{graphicx}

\RequirePackage[nomain,nonumberlist,nogroupskip]{glossaries}
\RequirePackage{xspace}
\RequirePackage{bm}
\RequirePackage{amsfonts}
\newglossary{symbols}{sym}{sbl}{List of Symbols}

\newcommand{\sym}[4]{\newglossaryentry{#1}{type=symbols,
        sort=#2,
        name={\ensuremath{#3}\xspace},
        description={#4}}%
    \expandafter\newcommand\expandafter{\csname #1\endcsname}{\gls{#1}}%
}

\newcommand{\symHidden}[4]{\newglossaryentry{#1}{type=hiddensymbols,
        sort=#2,
        name={\ensuremath{#3}\xspace},
        description={#4}}%
    \expandafter\newcommand\expandafter{\csname #1\endcsname}{\gls{#1}}%
}

\newcommand{\symLower}[6]{\newglossaryentry{#1#2}{type=symbols,
        sort=#3,
        text={\ensuremath{#4}\xspace},
        name={\ensuremath{#4_{#5}}\xspace},
        description={#6}}%
    \expandafter\newcommand\expandafter{\csname #1\endcsname}[1]{\ensuremath{\gls{#1#2}_{##1}}}%
    \expandafter\newcommand\expandafter{\csname #1#2\endcsname}{\ensuremath{\gls{#1#2}_{#5}}}%
}

\newcommand{\symLowerComma}[6]{\newglossaryentry{#1#2}{type=symbols,
        sort=#1#2,
        text={\ensuremath{#3}\xspace},
        name={\ensuremath{#3_{#4,#5}}\xspace},
        description={#6}}%
    \expandafter\newcommand\expandafter{\csname #1\endcsname}[1]{\gls{#1#2}_{#4,##1}}%
    \expandafter\newcommand\expandafter{\csname #1#2\endcsname}{\gls{#1#2}_{#4,#5}}%
}

\newcommand{\symFun}[6]{\newglossaryentry{#1#2}{type=symbols,
        sort=#3,
        text={\ensuremath{#4}\xspace},
        name={\ensuremath{#4\left({#5}\right)}\xspace},
        description={#6}}%
        \expandafter\newcommand\expandafter{\csname #1\endcsname}[1]{\gls{#1#2}\!\left({##1}\right)}%
        \expandafter\newcommand\expandafter{\csname #1#2\endcsname}{\gls{#1#2}\!\left({#5}\right)}%
}

\newcommand{\linkset}[2]{\left\{#1 \,|\, #2 \right\}}
\newcommand{\Prob}[1]{P\!\left(#1\right)}
\newcommand{\Exp}[1]{E\!\left[#1\right]}

\sym{Iup}{Iup}{I_{up}}{Mean packet generation interval}
\sym{pgenup}    {gnu}{g_{\text{up}}}{Packet generation rate in upstream}
\sym{poissonrate}    {lni}{\lambda_{n,i}}{Packet rate in the Poisson part of the traffic model}
\sym{bernoulliprob}    {b}{\beta_{n,i}}{Packet probability in the Bernoulli part of the traffic model}

\newcommand{\mtr}[2]{P\!\left(#1 \to #2\right)}
\newcommand{\qqtr}[2]{\mtr{\qstate{#1}}{\qstate{#2}}}

\symLower{qstate}{qi}        {sqi}{\sigma^n}{q,i}{State in the queuing model}

\symLowerComma{Paccept}{Pacceptn}{P}{\text{accept}}{n}{Packet accepting probability}

\symLower{Arr}{ni}{Ani}{\bm{A}}{n,i}{Random variable for the number of arriving, i.e. generated and received, packets in slot $i$ at node $n$}
\symLower{Atotal}{n}{A}{\mathcal{A}}{n}{Total expected number of generated or received packets per slotframe}
\symLower{Que}{n}{Qn}{\bm{Q}}{n}{Random variable for the number packets inserted into the queue}
\sym{Poi}{Pni}{\bm{\Pi}_{n,i}}{Random variable for the number of arriving packets according to the Poisson part of the traffic model}

\symLower{Delay}{n}         {Dn}{\bm{D}}{n}{Random variable for the queuing delay}

\sym{states}{Sn}{\bm{\Sigma}_{n}}{The set of states for the queuing model}

\sym{lenslot}{Ts}{T_{s}}{Duration of a slot}
\symLower{Rx}{ni}        {mRni}{\mu^{\textsc{RX}}}{n,i}{Probability of receiving a packet in slot $i$}
\symLower{Tx}{ni}        {mTni}{\mu^{\textsc{TX}}}{n,i}{Probability of sending a packet in slot $i$}

\sym{lensched}{lS}{l_{S}}{Overall number of slots in the slotframe}

\sym{Sctn}{S}{\left\vert\mathcal{T}_n\right\vert}{Number of transmission slots in the schedule for node $n$}

\symLower{txslots}{n}        {Tn}{\mathcal{T}}{n}{Transmission slots}
\symLower{rxslots}{n}        {Rn}{\mathcal{R}}{n}{Reception slots}

\symLowerComma{txslot}{tni}{t}{n}{i}{Transmission slot}
\symLowerComma{rxslot}{rni}{r}{n}{i}{Reception slot}

\symLower{csteady}{nqi}        {cnqi}{c}{n,q,i}{Probability of being in state $\qstate{q,i}$ in the stationary distribution}
\sym{cvect}{c}{c}{Vector of the $\csteady{n,q,i}$}

\sym{rootnode}{v0}{v_0}{Root node}
\symLower{parent}{n}        {pn}{p}{n}{Parent of node $n$}
\symLower{node}{n}        {vn}{v}{n}{Node with index $n$}

\sym{channels}{c}{\mathcal{C}}{Channels}
\symLower{neighbors}{n}        {Nn}{\text{\bf N}}{n}{Neighbors of node $n$}
\symLower{pdesc}{n}         {gn}{\gamma}{n}{Number of proper descendants of node $n$}
\symLower{counterpart}{n}        {Kn}{\mathcal{K}}{n}{Relation of slots to counterpart}
\symLower{channel}{n}        {Cn}{\mathcal{C}}{n}{Relation of slots to channel}
\symLower{blockedchannels}{n}        {Bn}{\mathcal{B}}{n}{Relation of slots to blocked channels}

\symLower{PER}{bl}{PERbl}{\text{PER}}{b,l}{Packet error rate for a transmission of $b$ bytes}
\sym{Li}		{Li}{{\bf L}_i}{The set of active links during slot $i$}
\symLower{Disturb}{il}{Dil}{{\bm{\mathcal{D}}}}{i,l}{The set of possibly disturbing links during slot $i$}
\symFun{inrange}{vw}{D}{\ensuremath{D}}{v,w}{Predicate that describes potential collisions between nodes}
\symLowerComma{Rup}{n}{\mathfrak{R}}{\text{up}}{n}{Reliability that a packet sent by $n$ arrives at the sink}
\symLowerComma{Dup}{n}{\mathfrak{D}}{\text{up}}{n}{End-to-end packet delay}

\makenoidxglossaries

\begin{document}

\begin{frontmatter}

\begin{fmbox}
\dochead{Research}


\title{An Analytical Model for Wireless Mesh Networks with Collision-Free TDMA and Finite Queues}


\author[
   addressref={aff1},                   
   corref={aff1},                       
   email={florian.kauer@tuhh.de}   
]{\fnm{Florian} \snm{Kauer}}
\author[
   addressref={aff1},
   email={turau@tuhh.de}
]{\fnm{Volker} \snm{Turau}}


\setattribute{authorinfo}       {text} {\endgraf}
\address[id=aff1]{%
  \orgname{Hamburg University of Technology, Institute of Telematics},
    \street{Am Schwarzenberg-Campus 3},
  \postcode{21073}
  \city{Hamburg},
  \cny{Germany}
}


\begin{abstractbox}

\begin{abstract} 
Wireless mesh networks are a promising technology for connecting sensors and actuators with high flexibility and low investment costs.
In industrial applications, however, reliability is essential.
Therefore, two time-slotted medium access methods, DSME and TSCH, were added to the IEEE 802.15.4 standard.
They allow collision-free communication in multi-hop networks and provide channel hopping for mitigating external interferences.
The slot schedule used in these networks is of high importance for the network performance. This paper supports the development of efficient schedules by providing an analytical model for the assessment of such schedules, focused on TSCH.
A Markov chain model for the finite queue on every node is introduced that takes the slot distribution into account. The models of all nodes are interconnected to calculate network metrics such as packet delivery ratio, end-to-end delay and throughput.
An evaluation compares the model with a simulation of the Orchestra schedule.
The model is applied to Orchestra as well as to two simple distributed scheduling algorithms to demonstrate the importance of traffic-awareness for achieving high throughput.
\end{abstract}



\end{abstractbox}
\end{fmbox}

\end{frontmatter}



\hyphenation{ma-the-ma-ti-cal}

\section{Introduction}

Wireless multi-hop networks are currently on the transition from a popular research topic to the application in real-world scenarios in the industry. Examples include oil refineries \cite{odonovan_ginseng_2013} and solar tower power plants \cite{pfahl_holistic_2014}. However, for such applications, the reliability of state-of-the-art techniques is not sufficient. This can often be traced back to packet collisions due to badly coordinated access to the wireless channel \cite{meiermodel}. Therefore, the IEEE has recently extended the IEEE~802.15.4 standard with two approaches for time-division multiple access (TDMA) in multi-hop networks, namely DSME and TSCH \cite{802154}.
Especially in large networks under heavy load, these approaches are expected to perform much better than the commonly used CSMA/CA.
While several implementations of these extensions already exist \cite{towards_openDSME,orchestra},
evaluations of the theoretical performance boundaries have not found much attention in the literature.
This paper closes the gap by providing a framework for analytical evaluation of such networks.

The performance evaluation of techniques for wireless networks is possible in multiple ways. Building real-world testbeds provides good insights into the performance of an actual application. However, it is very cost- and time-intensive. 
Therefore, event-based simulators are developed that try to replicate the reality as closely as possible, while allowing reproducible experiments without the need of specialized hardware.
The disadvantage is a complex and extensive implementation
and large execution times.

An alternative is the analysis of mathematical models. An analytical model in the sense of this paper is a system of equations that can be solved by numerical means. While such a model usually only describes a small subset of the behavior of a real system and is therefore only an approximation to the real behavior, the calculations are often faster than a simulation by magnitudes and influences from external sources are avoided. The results can also be verified and reproduced quite easily.
Last but not least, the comparison of a simulator and an analytical model helps to find inaccuracies and bugs in both approaches and gives new insights into the underlying principles.

The main contributions of the paper are as follows:
\begin{itemize}
\item A Markov chain model for a node's finite queue that considers the slot schedule.
\item A multi-hop model for calculating packet delivery ratio, end-to-end delay and throughput. 
\item An open-source implementation of the models \cite{modelimpl}.
\item An extensive evaluation, including a comparison with an event-based simulation and the application to two traffic-aware schedules for TSCH.
\end{itemize}

After presenting the related work in the next section, the requirements of a model for TSCH are outlined in Sect.~\ref{sect:requirements} followed by the system model in Sect.~\ref{sect:systemmodel}. We then take the angle of a single node and present and evaluate the queuing model in Sect.~\ref{sect:queuemodel}. These models are then concatenated in Sect.~\ref{sect:multihop} to derive global network metrics and a comparison with a COOJA simulation is presented. In Sect.~\ref{sect:schedules}, two algorithms for building slot schedules are introduced, evaluated and compared to the Orchestra SBD schedule. Sect.~\ref{sect:extensions} gives an outlook to possible extensions and the paper is finally concluded in Sect.~\ref{sect:conclusion}.

\section{Related Work}
Analyzing the performance of wireless networks is a challenging research topic of high value for practical applications \cite{pathak_survey_2011}.
Many evaluations for IEEE 802.15.4 networks are based on simulations \cite{zheng_comprehensive_2006} or real-world testbeds \cite{dutta_trio_2006}.
Especially for the CSMA/CA technique of IEEE 802.15.4 analytical models exist for single-hop topologies \cite{misic_performance_2006} and multi-hop networks \cite{dimarco_analytical_2012,meiermodel}. In these publications, packet collisions, especially due to hidden node constellations, are identified as a major reason for bad performance of CSMA/CA.

Since their admission to the standard, TSCH and DSME were analyzed by simulations \cite{alderisi_simulative_2015}, testbeds \cite{orchestra} and analytical models. The latter includes work focused on special aspects such as network formation \cite{guglielmo_performance_2014}
or transmission in the contention access period (CAP) \cite{sahoo_novel_2017}. The latter is very similar to the analysis of conventional CSMA/CA because it does not take guaranteed time slots into account.
A simple formula for the throughput of DSME networks is given in \cite{jeong_performance_2012}.
Furthermore, much research exists that analyzes TDMA communication on a more general level \cite{gronkvist_throughput_2004,bjorklund_resource_2003}.

A major aspect of network analysis is the influence of the queue filling levels. The fundamentals of queuing theory are well-established \cite{ng_queueing_2008}.
Especially the M/D/1/K model is of particular interest for TDMA networks, describing Poisson distributed arrival, deterministic service time, a single transmitter and a queue of length $K$. In \cite{seo_explicit_2014} closed-form expressions are given for the blocking probability and other properties. 
Many variants were analyzed including service times following a general distribution \cite{macgregor_smith_properties_2011}. 
In \cite{khan_delay_1998} a generalized queuing model for TDMA is developed that is also applicable for traffic that does not follow a Poisson distribution. 

In this paper, a queuing model is presented that respects the specific structure of dedicated schedules, in particular, but not limited to, IEEE 802.15.4 TSCH networks. A popular schedule implemented in Contiki \cite{dunkels_contiki_2004} is Orchestra \cite{orchestra}. It does not require any management traffic for building up the schedule. Part of the ongoing IETF standardization of 6TiSCH, the link between IPv6 networks and TSCH, is the development of scheduling functions such as SF0 \cite{dujovne_6tisch_2017_sf0} and SF1 \cite{anamalamudi_scheduling_2017}.

Many other scheduling techniques were proposed that are applicable to TSCH networks or to one of its forerunners, the WirelessHART standard \cite{wirelesshart}. A centralized schedule for the latter is presented in \cite{pottner_constructing_2014}. 
Another centralized schedule, this time for TSCH, is the Traffic Aware Scheduling Algorithm (TASA) \cite{palattella_optimal_2013}.
Traffic awareness is an important property for improving throughput as we will also see in the evaluation of this paper. However, centralized scheduling algorithms come with a high overhead for building, distributing and maintaining the schedule, especially in large networks. Therefore, many decentralized approaches were suggested, such as DeTAS \cite{accettura_decentralized_2015} 
and DIS\_TSCH \cite{hwang_distributed_2017}.

Another promising distributed algorithm to generate a conflict-free schedule is Wave, presented in \cite{soua_wave_2016} and extensively analyzed in \cite{soua_wave_2016_ext}. 
DeBraS \cite{municio_decentralized_2016} explicitly targets dense networks where colliding slots are a major problem.
All mentioned algorithms implicitly minimize energy consumption by reducing the number of slots assigned to every node. In contrast, \cite{ojo_energy_2017} 
explicitly targets minimum energy consumption by formulating an energy efficiency maximization problem. 

\section{Requirements}
\label{sect:requirements}
The goal of the paper is to provide a tool for estimating the achievable performance of a network using the IEEE 802.15.4 TSCH data link layer with given multi-hop topology and associated schedule. In this section, an analysis of the potential sources for packet loss is presented. These are compared to the features provided by modern TDMA medium access protocols such as IEEE~802.15.4 TSCH and DSME. The most relevant aspects for a model are finally summarized.

The main reasons for packet loss in wireless multi-hop networks can be coarsely categorized as follows.
\begin{itemize}
\item Collisions of transmissions of the same network.
\item External interferences, for example between IEEE 802.15.4 and IEEE 802.11 transceivers or with another unsynchronized IEEE 802.15.4 network.
\item High path loss, for example due to a large distance or fading.
\item Queue drops if the rate of packets that are generated or have to be forwarded is higher than the rate at which packets can be transmitted. This can be permanent or during a burst.
\end{itemize}

To mitigate these losses, the introduced TDMA data link layers TSCH and DSME provide the following features that allow for reliable multi-hop communication.

\begin{itemize}
\item A slot structure with predefined timing to arrange the transmissions in the time domain to avoid collisions within the same network. 
\item Time synchronization to ensure an aligned timing throughout the multi-hop network. 
\item Channel adaption (only DSME) and channel hopping (both) to arrange transmission in the frequency domain to increase the number of transmissions per time as well as to mitigate external interferences.
\item Procedures for setting up time and frequency schedules in a distributed fashion to ensure conflict-free transmissions. This is only integrated in DSME, but the ongoing IETF standardization of 6TiSCH provides related features for TSCH.
\end{itemize}

To calculate the performance in steady-state, it can be assumed that all nodes are already associated to the network, are properly synchronized to the global notion of time, and have negotiated a conflict-free and valid multi-hop schedule. In such a schedule, a combination of a time slot and a frequency channel is either free or assigned to exactly one transmitter and one receiver in every neighborhood. By this, packet collisions can be avoided, even in hidden node constellations. In DSME networks this constraint is inherent, given that no failures happen during the negotiation \cite{formaldsme}. For TSCH networks, such a fixed assignment is not necessarily required, but sharing will obviously lead to collisions, even if they might not be as severe as with CSMA/CA. Furthermore, the schedule should avoid using links with high path loss or highly fluctuating channel conditions to avoid occasional packet loss on the physical layer by means of an appropriate neighborhood management \cite{Telematik_Adhoc-Now_2015}.

In large industrial plants, the frequencies of the used radio components are usually coordinated and monitored to avoid cross-interferences and as a security measure \cite{wirelessnerc}. However, external interferences can often not be avoided completely, so TSCH and DSME can dynamically use multiple frequency channels. Apart from mitigating external interferences, this can also be used to increase the throughput by assigning the same time slot to multiple transceiver pairs on different frequency channel and is therefore also modeled in this paper. It can either be implemented as channel adaption where a fixed frequency channel is assigned at a given time or channel hopping where the channel to be used is iterated over a given sequence. Yet, in this paper external interferences are not considered, so channel hopping with non-overlapping hopping sequences is equivalent to a fixed assignment as for channel adaption.

It is therefore concluded that in properly constructed TSCH and DSME networks without a lot of external disturbances, the occurrence of queue drops is the main factor that determines the maximum achievable network performance. Yet, in real-world networks other losses can still occur, so Sect.~\ref{sect:otherloss} gives an outlook about how to consider these in the model.

As noted in the previous section, the M/D/1/K model is often used to model queues in the context of TDMA networks. However, it has two major weaknesses in the given scenario: Even if the traffic generation itself is modeled as Poisson distributed, this distribution does not necessarily hold for forwarding nodes, so the M/D/1/K model is not suitable for multi-hop networks. Furthermore, the service times are not necessarily deterministic but can diverge significantly, due to two effects: First, the schedule itself might be irregular, for example if multiple subsequent transmission slots are followed by a large idle phase. Secondly, even if the schedule is regular, the service time of packet that arrives at an empty queue is not constant, but depends on the time left until the next transmission slot. This effect is especially relevant in scenarios with low traffic. A proper model must take these effects into account.

\section{System Model} 
\label{sect:systemmodel}
\begin{figure}[h]%
\centering\includegraphics{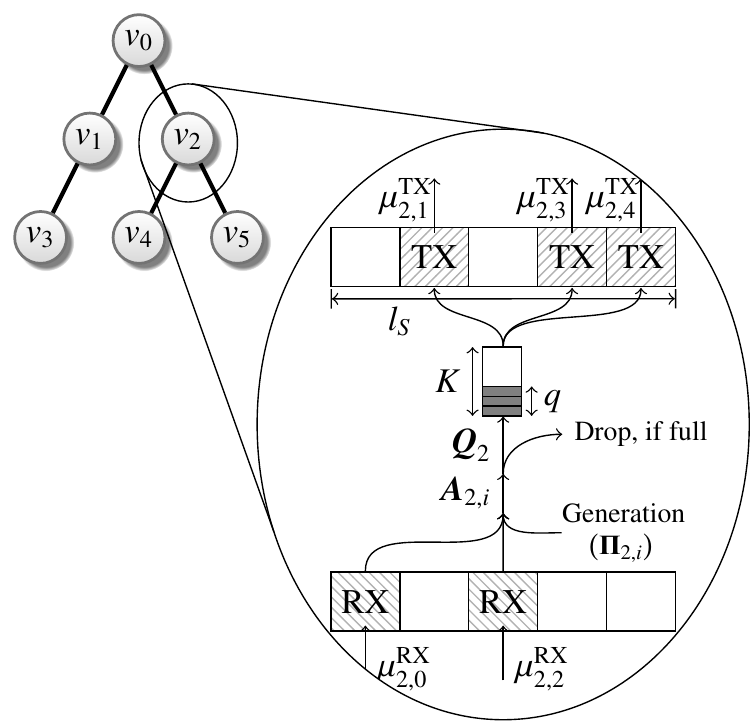}%
  \caption{Illustration of the proposed model annotated with the notation introduced in the following sections.\label{fig:notationoverview}}
\end{figure}

In the presented system model as illustrated in Fig.~\ref{fig:notationoverview} the queue of every node is modeled as an instance of a Markov chain and they are linked to form a model of the full network. Packets are either forwarded as received or generated at the node. In both cases, the packets are pushed to a queue of fixed length or dropped if the queue is full. If there is at least one packet in the queue at the beginning of a transmission slot, a transmission attempt takes place.

The calculations are based on fixed multi-hop schedules. A schedule can be calculated offline with algorithms such as the ones presented in Sect.~\ref{sect:schedules}. Alternatively, a schedule can be extracted from a real-world deployment or a simulator as presented in Sect.~\ref{sect:simulation}. In the following, the notation for such a schedule as used in the model is introduced.

\begin{figure}[h]%
\centering\includegraphics{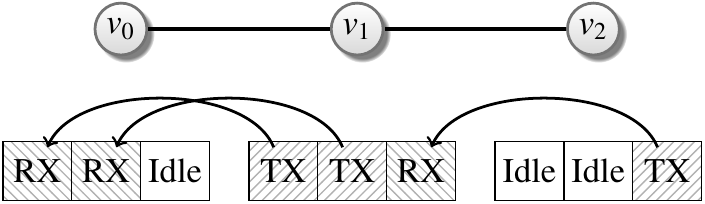}%
    \caption{An exemplary schedule corresponding to $N=3$, $\lensched=3$, $\txslots{0} = ()$, $\rxslots{0}=(0,1)$, $\counterpart{0}(0)=1$, $\counterpart{0}(1)=1$, $\txslots{1}=(0,1)$, $\rxslots{1}=(2)$, $\counterpart{1}(0)=0$, $\counterpart{1}(1)=0$, $\counterpart{1}(2)=2$, $\txslots{2}=(2)$, $\rxslots{2}=()$ and $\counterpart{2}(2)=1$.\label{fig:example_schedule}}
\end{figure}

For the presented model, a network of $N$ nodes $\node{0},\ldots,\node{N-1}$ with a fixed schedule is considered. All slots have an equal time duration $\lenslot$. For TSCH it is usually $10\,\text{ms}$. A slotframe consists of $\lensched$ slots and is repeated with an interval of $\lenslot\cdot\lensched$.  The schedule for node $\node{n}$ is given as a tuple of transmission slots $\txslots{n} = (\txslot{j})_{j=0,\ldots,\Sctn-1}$, sorted in ascending order, where $\txslot{j} \in \txslots{n}$ if and only if the slot with the zero-based index $\txslot{j}$, that is $0 \leq \txslot{j} < \lensched$, is assigned to $n$ for conflict-free transmission. Correspondingly, $\rxslots{n} = (\rxslot{j})_{j=0,\ldots,\left\vert\rxslots{n}\right\vert-1}$ describes the reception slots. A slot is never in both $\txslots{n}$ and $\rxslots{n}$, because a node can not transmit and receive at the same time. Furthermore, $\counterpartn(i)$ gives the respective reception node if $i \in \txslots{n}$ or the respective transmission node if $i \in \rxslots{n}$. Fig.~\ref{fig:example_schedule} presents a simple example for these definitions.

For modeling frequency diversity, $\channel{\node{n}}(i)$ is introduced that specifies the channel to use in slot $i$ by node $\node{n}$. The channel is an element of the set of channels $\channels$. For IEEE 802.15.4 in the $2.4\,\text{GHz}$ band, it consists of the numbers $11$ to $26$.

To check that a schedule is conflict-free and valid, the set
\begin{align}
  \Li = \left\{(n,\counterpart{n}(i))\,\middle|\, i \in \txslots{n}, \forall n\right\}
\end{align}
of all links that are active during slot $i$ has to be considered for all slots.
Each node can only be transmitter \emph{or} receiver during a slot
\begin{align}
  \left(i \in \txslots{n} \Rightarrow i \notin \rxslots{n}\right) \wedge \left(i \in \rxslots{n} \Rightarrow i \notin \txslots{n}\right).
\end{align}
Furthermore, the links have to be unique and consistent, $\forall i \in \txslots{n}$ it must hold
\begin{align}
  i \in \rxslots{\counterpart{n}(i)} \wedge \left(\counterpart{n}(i)=k \Leftrightarrow \counterpart{k}(i)=n\right).
\end{align}

\begin{figure}[tb]
\centering\includegraphics{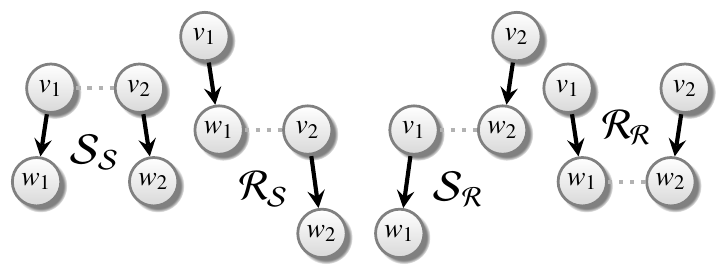}
\caption{Possible conflicts between two transmissions.\label{fig:rel}}
\end{figure}

However, this alone is insufficient. Other potential conflicts are shown in Fig.~\ref{fig:rel}. The most obvious constellation is $\mathcal{R_S}$: The reception at $w_1$ can be disturbed by a transmission from $v_2$ to $w_2$ if $v_2$ is in the neighborhood of $w_1$. If acknowledgments are used, conflicts between acknowledgments and the data packets have to be taken into account, too, represented by the other constellations. For a link $(v_1,w_1) \in \Li$ the potentially disturbing links are
\begin{align}
  \begin{aligned}
      &\hspace{-0.3cm} \Disturb{i,(v_2,w_2)}= \\
    & \linkset{(v_2,w_2) \in \Li}{\inrange{v_1,v_2} \wedge v_1 \neq v_2}\\
  &\cup \linkset{(v_2,w_2) \in \Li}{\inrange{w_1,v_2} \wedge v_1 \neq v_2}\\
 &\cup \linkset{(v_2,w_2) \in \Li}{\inrange{v_1,w_2} \wedge v_1 \neq v_2}\\
 &\cup \linkset{(v_2,w_2) \in \Li}{\inrange{w_1,w_2} \wedge v_1 \neq v_2}.
  \end{aligned}
\end{align}
Here $\inrange{v,w}$ denotes that a transmission of $v$ might disturb a reception at $w$. One possibility to ensure a conflict-free schedule is to have 
\begin{align}
  \Disturb{i,\cup} = \bigcup_{l\,\in\,\Li} \Disturb{i,l}
\end{align}
empty for all slots. Hence, no concurrent transmissions take place in the neighborhood. By exploiting frequency diversity it is, however, sufficient to make sure that no concurrent transmissions take place on the same frequency channel at the same time, i.e. one has to make sure that for all slots $0 \leq i < \lensched$ holds
\begin{align}
  \channel{v_1}(i) \neq \channel{v_2}(i), \forall (v_2,w_2) \in \Disturb{i,\left(v_1,w_1\right)}
\end{align}

\section{Queue Model}
\label{sect:queuemodel}

Before modeling multi-hop communication, this section considers the viewpoint of a single node $\node{n}$ and its local behavior. In the multi-hop model, every node of the network will get its own instance of this model. The main focus is the finite queue, so after this section we will be able to calculate packet drops due to a full queue as well as the expected queuing delay. The inputs into this model are the probabilities for incoming traffic and the distribution of the reception slots $\rxslots{n}$ and transmission slots $\txslots{n}$. Beyond the usage in a multi-hop model as introduced in the next section, the model in this section is also applicable in other setups, such as single-hop networks or for evaluating the performance of a single node as presented in Sect.~\ref{sect:singlenodeeval}.

The following policy is used to model the queue:

\begin{itemize}
\item The queue can hold at most $K$ packets.
\item The number of packets in the queue at the beginning of a slot is denoted as $q$.
\item New packets can arrive at any time during a slot. At most $K-q$ packets are accepted during a slot.
\item A packet is removed from the queue at the end of a slot if and only if it is a transmission slot and the packet was already in the queue at the beginning of the slot, i.e. $q > 0$, modeling the transmission process itself.
\end{itemize}

\begin{figure}[t]%
\centering\includegraphics{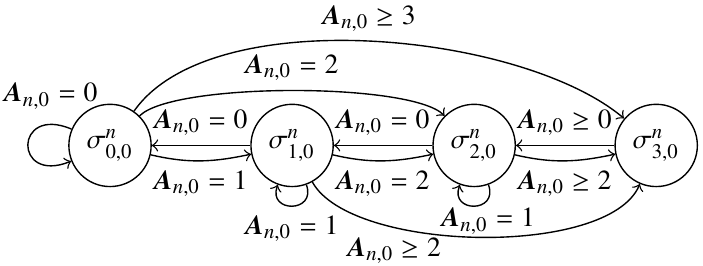}%
    \caption{A M/D/1/K model in the syntax introduced in this paper with $K = 3$, $\lensched = 1$ and $\txslotsn= (0)$.\label{fig:md1k}}
\end{figure}

\begin{figure*}[t]%
\centering\includegraphics{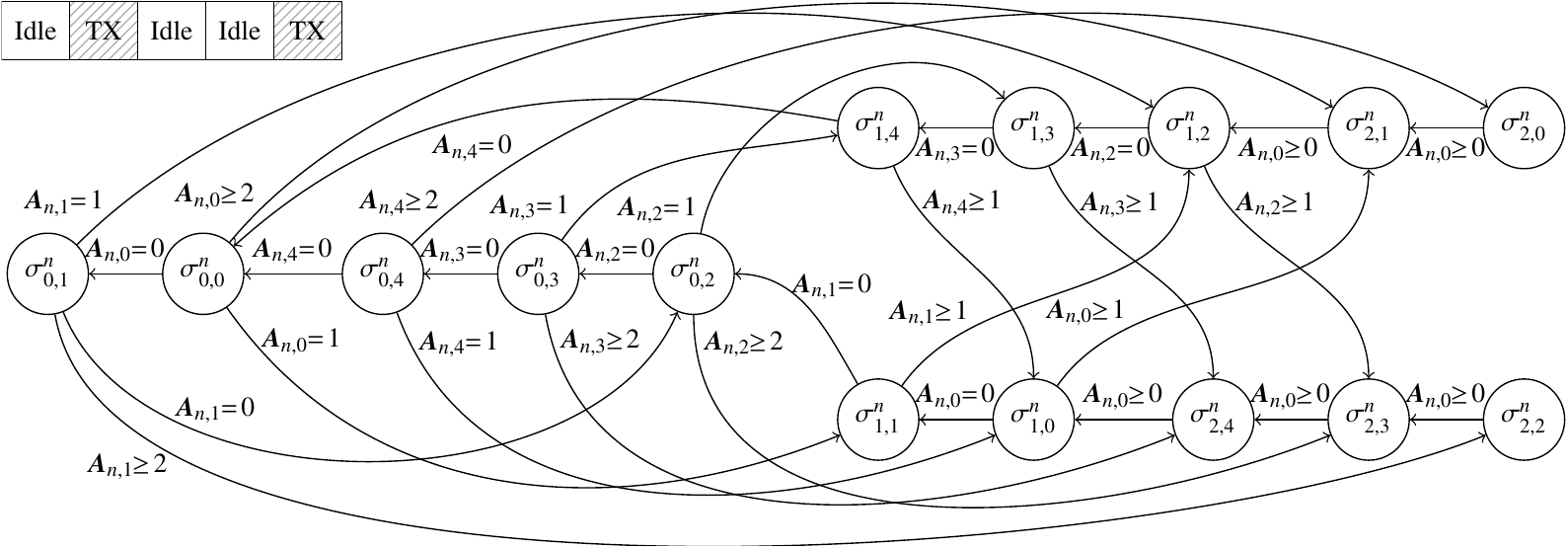}%
    \caption{Schedule for $\lensched = 5$, $\txslotsn=(1,4)$ and corresponding state diagram for $K=2$.\label{fig:schedule}}
\end{figure*}

The queue of a node is modeled as a discrete-time Markov chain with the states
\begin{align}
\states = \left\{ \qstate{q,i} \,\middle|\, 0 \leq q \leq K, 0 \leq i < \lensched \right\}.
\end{align}
In order to account for the irregular schedule and for modeling the service time more accurate, the state of the queue $\qstate{q,i}$ does not only account for the number $q$ of packets in the queue as for usual M/D/1/K models, but also the current position within the slot schedule $i$. As shown in Fig.~\ref{fig:md1k}, the model presented here is a super set of the M/D/1/K model for $\lensched=1$ and $\txslotsn=(0)$.
Fig.~\ref{fig:schedule} depicts a more complex schedule, the corresponding states and the transitions as follows.


\subsection{Traffic Model}
\label{sect:traffic}

The number of arriving packets during a time slot $i$, that is the generated and received traffic, is described by the random variable $\Arr{n,i}$. For the purpose of the queuing model it can follow an arbitrary distribution that can optionally depend on the current state. A popular option is to model the traffic as Poisson distributed with a mean packet rate $\poissonrate$. For this, the random variable $\Poi$ is introduced with the probability distribution
\begin{align}%
\Prob{\Poi = k} = \frac{\poissonrate^k}{k!} e^{-\poissonrate}.
\end{align}
The probability for \emph{at least} $k$ packets is given by
\begin{align}%
\Prob{\Poi \geq k} = 1 - \sum_{j=0}^{k-1} \Prob{\Poi = k}.
\end{align}

Since only a single packet can be received per slot, we extend this model by combining it with a Bernoulli distribution with the probability $\bernoulliprob$ that a single packet is arriving in addition to the Poisson traffic in slot $i$, so we get
\begin{multline}%
    \Prob{\Arr{n,i} = k} =\\
  \begin{cases}
      \!\!\begin{array}{l}\left(1\!-\!\bernoulliprob\right)\cdot\Prob{\Poi = 0}\end{array} & k = 0\\
      \!\!\begin{array}{l}\left(1\!-\!\bernoulliprob\right)\cdot\Prob{\Poi = k}\\
      \hspace{0.3cm}+ \,\bernoulliprob\cdot\Prob{\Poi = k-1}
      \end{array}\!\!\!\!\! & \text{otherwise}
  \end{cases}
\end{multline}
and
\begin{multline}%
    \Prob{\Arr{n,i} \geq k} =\\
  \begin{cases}
      \!\!\begin{array}{l}1\end{array} & k = 0\\
      \!\!\begin{array}{l}\left(1\!-\!\bernoulliprob\right)\cdot\Prob{\Poi \geq k}\\
      \hspace{0.3cm}+ \,\bernoulliprob\cdot\Prob{\Poi \geq k-1}
      \end{array}\!\!\!\!\!\!\!\! & \text{otherwise.}
  \end{cases}
\end{multline}

Since all $\lensched$ slots occur with the same probability, the expected total number of packets per slotframe is calculated as the sum of the 
expected number of the $\Arr{n,i}$ as derived in Appendix~A as
\begin{align}
    \begin{aligned}
    \Atotal{n} &= \sum_{i=0}^{\lensched-1} \Exp{\Arr{n,i}}\\
    &= \sum_{i=0}^{\lensched-1}\left(1-\bernoulliprob\right) \cdot \poissonrate + \bernoulliprob \cdot \left(\poissonrate + 1\right).
    \end{aligned}
\end{align}%

\subsection{Queuing Probability}
The random variable $\Que{n}$ is the number of accepted packets that are inserted into the queue per slot. Since at most $K-q$ packets can be inserted into the queue, its probability distribution, i.e. the probabilities of queuing $k$ packets, is calculated as
\begin{multline}%
    \Prob{\Que{n} = k \,\middle|\, \qstate{q,i}} =\\ \begin{cases}
      \Prob{\Arr{n,i} = k} & k < K - q \\
        \Prob{\Arr{n,i} \geq K-q} & k = K - q\\
        0 & \text{otherwise.} \\
\end{cases}
\end{multline}
Obviously, the difference of $\Arr{n,i}$ and $\Que{n}$ is the number of packets dropped due to a full queue.

\subsection{Transition Probabilities}
$\mtr{\xi}{\zeta}$ with $\xi,\zeta \in \states$ is the probability of going from $\xi$ to $\zeta$ in one step. Transition probabilities not listed are zero, in particular for transitions with non-consecutive slots.
Furthermore, $\%$ denotes the modulo operation. So for $0 \leq q \leq K$ and $0 \leq i < \lensched$, all possible transitions are specified by%
\begin{multline}%
  \qqtr{q,i}{\max\left(q-\tau_{n,i},0\right)+k,\,\,(i+1) \% \lensched}\\ = \Prob{\Que{n} = k \,\middle|\, \qstate{q,i}}
\end{multline}
where%
\begin{align}%
  \tau_{n,i} = \begin{cases}
        1 & i \in \txslots{n}\\
        0 & i \not\in \txslots{n}
    \end{cases}
\end{align}
and $k$ is the number of arriving or newly generated packets during the time slot that iterates from $0$ to $K-q$.
While this gives a complete description of the model, two corner cases should be mentioned explicitly for clarity. With an empty queue and $\Arr{n,i}=0$, that is no arriving packets, the model stays in the states with $q=0$. With a full queue, it is not possible to insert further packets into the queue so $\Que{n} = 0$ and thus the single possible transition out of the states with $q = K$ has the probability
\begin{align}
\Prob{\Que{n} = 0 \,\middle|\, \qstate{K,i}} = \Prob{\Arr{n,i} \geq 0} = 1,
\end{align}
either to a state with $q = K-1$ if the current slot is a transmission slot or else to a state with $q = K$.

As shown in Appendix B, these probabilities can be used to calculate the stationary distribution of this Markov chain. 
This results in the probability $\csteady{n,q,i}$ of being in state $\qstate{q,i}$ in the stationary case.

\subsection{Transmission Probability}

From the $\csteady{n,q,i}$, the probability that a successful transmission takes place within a given slot is calculated from the complementary probability of being in a state with empty queue ($q=0$) as%
\begin{align}
\Tx{n,i} = \tau_{n,i} \cdot \left(1 - \frac{\csteady{n,0,i}}{\sum_{q=0}^{K} \csteady{n,q,i}}\right).
\end{align}

\subsection{Packet Acceptance Probability}
\label{sect:paccept}

If $q$ packets are in the queue, $K-q$ packets can be inserted into the queue, so the expected number of \emph{accepted} packets in state $\qstate{q,i}$ is
\begin{align}
\begin{aligned}
    \Exp{\Que{n} \,\middle|\, \qstate{q,i}} \!=\! \,\,&\Prob{\Arr{n,i}\!\geq\!K-q } \cdot (K-q) \\
                                         &+ \sum_{k=0}^{K-q-1} \Prob{\Arr{n,i} = k}\cdot k.
\end{aligned}
\end{align}
Since the events of being in a state are mutually exclusive and exhaustive, $\Exp{\Que{n}}$ can be calculated from this according to the law of total expectation as
\begin{align}
    \Exp{\Que{n}} = \sum_{q=0}^{K} \sum_{i=0}^{\lensched-1} \csteady{n,q,i} \cdot \Exp{\Que{n} \,\middle|\, \qstate{q,i}}
\end{align}
and the overall expected number of packets per slotframe is thus $\lensched\cdot\Exp{\Que{n}}$. Together with the overall expected number of packets arriving at the queue, the overall packet acceptance probability is calculated as%
\begin{align}
    \Paccept{n} = \frac{\lensched\cdot\Exp{\Que{n}}}{\Atotal{n}}.
\end{align}

\subsection{Queuing Delay}
The number of time steps it takes until an arriving packet is transmitted is described by the random variable $\Delay{n}$. For convenience, we define%
\begin{align}
    \phi_{n}(i) = \begin{cases}
        \Sctn-1 & i \leq \txslot{0} \,\,\vee\\
         & i > \txslot{\Sctn-1}\\
        g : \txslot{g} < i \leq \txslot{g+1}\!\!\! & \text{else,}
    \end{cases}
\end{align}
as the index of the transmission slot preceding $i$ and%
\begin{align}
    \delta\!\left(i,j\right) = \begin{cases}
j - i & j \geq i\\
j - i + \lensched & \text{otherwise,}
\end{cases}
\end{align}
as the number of slots when going from $i$ to $j$. Then, given the event of being in state $\qstate{q,i}$ after accepting a packet and pushing it in the queue, $\Delay{n}$ is calculated as
\begin{align}%
    \Delay{n}\!\left(\qstate{q,i}\right) = &f_q \lensched + 1 + \delta\!\left(i,\,\,\txslot{\left(\phi_{n}(i) + q\right) \% \Sctn}\right)
\end{align}%
where $f_q = \left\lceil\tfrac{q}{\Sctn}-1\right\rceil$ is the number of iterations over the full slotframe, the summand $1$ accounts for the transmission slot itself and the last summand for the remaining packets after $f_q\cdot\Sctn$ packets were transmitted. The expected delay for a packet arriving at an arbitrary point in time can be calculated from this as%
\begin{align}%
  \begin{aligned}
  \Exp{\Delay{n}} &= \sum_{q=0}^{K}\sum_{i=0}^{\lensched-1} \csteady{n,q,i} \cdot \Delay{n}\!\left(\qstate{g,h}\right)\\
  \text{with }\qstate{g,h} &= \qstate{\max\left(q-\tau_n,0\right)+1,\,\,(i+1) \% \lensched}
  \end{aligned}
\end{align}
considering the additional transition to account for the arriving packet itself.

\subsection{Single Node Evaluation}
\label{sect:singlenodeeval}
\begin{figure}[bt]%
\centering\includegraphics{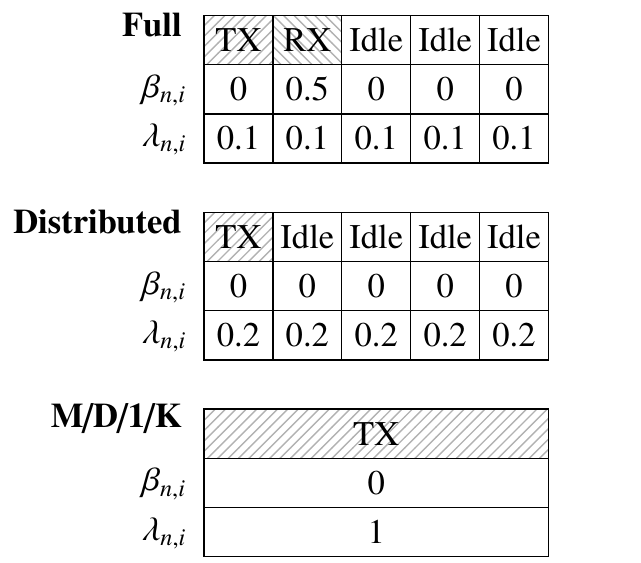}%
    \caption{Exemplary schedule for evaluating the queuing model together with two approximations, all with the same $\Atotal{n} = 1$. \label{fig:singlenode}}
\end{figure}
\begin{figure}[h]%
\centering\includegraphics{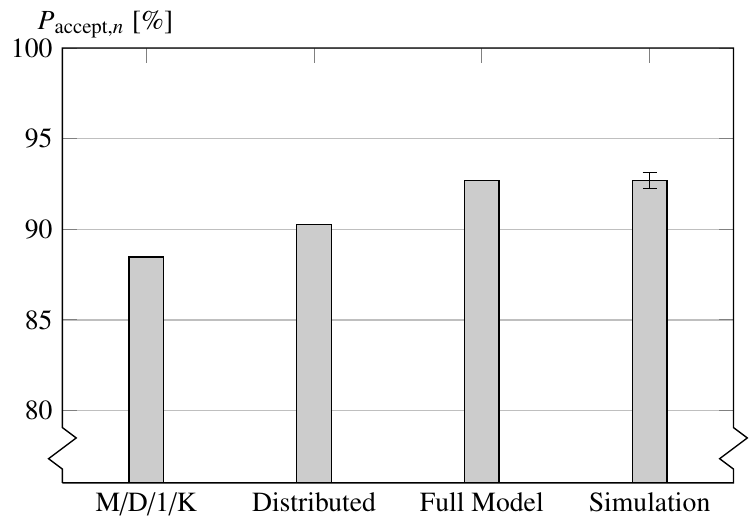}%
    \caption{Comparing the full presented model with a simulation and simpler models: An M/D/1/K model and a model that considers the TX and Idle slots, but distributes the reception over all slots.\label{fig:singlenodeeval}}
\end{figure}

 %
At first, the results for the queuing model on a single node are analyzed without considering the effects of multi-hop networks. The purpose of this section is to demonstrate the benefit of the refined model in contrast to a simple M/D/1/K model by comparing them with an event-based simulation. For this, an exemplary schedule is shown in Fig.~\ref{fig:singlenode} together with the two approximations presented in the following. The schedule is construed in a way that on average the same amount of traffic enters and leaves the system. However, due to a finite queue length of $K=5$, packet drops can still occur.
In Fig.~\ref{fig:singlenodeeval} the following four evaluations are compared:
\begin{itemize}
    \item In the \textbf{M/D/1/K} model, there is no option to specify multiple slots and only an overall packet rate rate can be specified. 
    \item In the \textbf{Distributed} approximation, the transmission and idle slots are modeled correctly, but incoming traffic is modeled to follow a Poisson distribution and not a Bernoulli distribution. Furthermore, it is distributed over all slots. To get the same $\Atotal{n}$ of 1 as in the actual scenario, all $\lambda_{n,i}$ are set to $\tfrac{1}{\lensched} = 0.2$. This scenario allows to assess the deficiency of not including the $\bernoulliprob$ in the model.
    \item The \textbf{Full Model} scenario additionally models the reception slot correctly as presented.
    \item The \textbf{Simulation} is a discrete event simulation of the queue based on SimPy \cite{simpy} that implements the policy as presented at the beginning of Sect.~\ref{sect:queuemodel}. The mean and $95\%$ confidence interval are shown of 10 runs with 10000 packets.
\end{itemize}

While the full model matches the simulation very well, the probability of dropping packets due to a full queue is higher in the distributed scenario, since the variance of the packet reception is higher, even though the average reception rate is equal. In fact, for $\pgenup = 0$ and at least as much TX slots as RX slots, the full model will never indicate a packet loss ($\Paccept{n} = 100\%$), while in the simplified models there is a non-zero probability that more packets are received in a slotframe than sent, eventually leading to packet loss.

The M/D/1/K model gives even less accurate results, because the service time is constant and equal to $\lensched\cdot\lenslot$, while in the distributed and the full model case, packets might be processed faster if the queue is empty and they arrive during the later slots.

\subsection{Queue Distribution}
\label{sect:queuedist}
In this section, we take a deeper look into the distribution of the queue level and its influence on the probability of accepting a packet for different traffic loads.
For this, a scenario with $K=10$, $\lensched=5$ and a single transmission slot is considered. 

\begin{figure}[htb]%
\centering%
\includegraphics{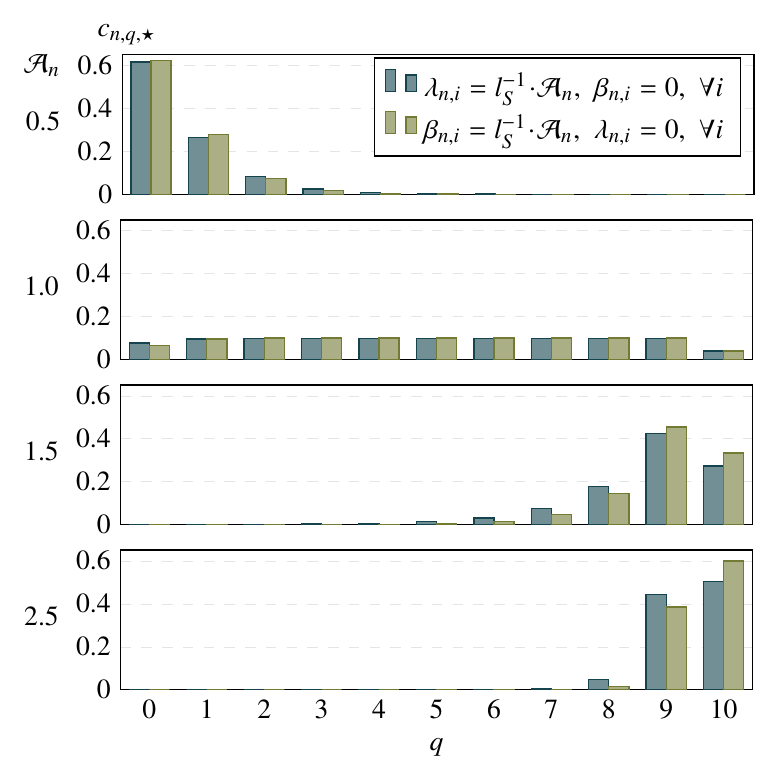}\\
    \caption{Distribution of queue levels for $K=10$, $\lensched=5$ and a single transmission slot.\label{fig:singlenodequeue}}
\end{figure}

In Fig.~\ref{fig:singlenodequeue}, the probability distribution of the queue level is plotted for different traffic loads $\Atotal{n}$. For this, the probability of being in a state with $q$ packets in the queue is calculated as%
\begin{align}%
\csteady{n,q,\star} = \sum_{i=0}^{\lensched-1} \csteady{n,q,i}.
\end{align}

In the first scenario, the traffic load is distributed over the $\poissonrate$, corresponding to a node that does only generate but does not forward traffic and in the second scenario, it is distributed only over the $\bernoulliprob$, corresponding to a node that only forwards traffic generated by other nodes.

For a low rate of $\Atotaln=0.5$, the queue is empty most of the time with a decaying probability of having more packets in the queue. The probabilities for queue levels larger than $3$ are negligible, so the probability of accepting new packets is $\Paccept{n}=1.00$ for both scenarios.

For a medium rate of $\Atotaln=1$, where on average one packet is sent and received every slotframe, the distribution is nearly constant. This leads to a packet acceptance probability of $\Paccept{n}=0.95$ for the first and $\Paccept{n}=0.96$ for the second scenario. Interestingly, the probability for $q=K$ is significantly smaller than the others.

This is due to the policy that in a transmission slot with $q=K$ no new packets can be pushed to the queue, but certainly a packet will be sent, i.e. $q=K-1$ in the next slot. However, for all other queue levels, it is possible to maintain the same queue level by sending and receiving one packet. That is even more apparent for $\Atotaln=1.5$ where a queue level larger than $q=5$ most of the time, leads to a packet acceptance probability of $\Paccept{n}=0.67$ for both scenarios. Having $q=K$ more often than other queue levels is only possible for very high rates of $\Atotaln = 2.5$. Here, the node is highly congested and a $\Paccept{n}=0.40$ is achieved in both scenarios.

When comparing the scenarios, we see that the difference is not very large, but in the second scenario, the probabilities are shifted to the boundaries. This evaluation also shows that often only a small part of the queue is actually utilized. For $\Atotaln = 0.5$ a maximum queue size of $K=3$ would be sufficient to prevent packet loss. Also for $\Atotaln = 2.5$, this would not change the packet acceptance probability. The probability distribution would shift to the left, but does not change its overall shape relative to the upper bound. Of course, the queuing delay would be reduced by this. The bottom line is that the maximum queue length $K$ has the highest influence in scenarios where the amount of incoming and outgoing traffic is similar.

\section{Multi-Hop Model}
\label{sect:multihop}

To model multi-hop communication, every node in the network gets its own instance of the previously presented model. They are then linked to get a full network for allowing to determine network-wide metrics.

\subsection{Traffic Generation and Forwarding}
For the purpose of this paper, we assume a data-collection scenario with sink $\rootnode$. Every node, except $\rootnode$, generates packets with exponentially distributed intervals with mean $\Iup$ that are to be forwarded to $\rootnode$ via a routing tree. The base time unit is the slot length $\lenslot$, so the generation rate is%
\begin{align}
\pgenup = \frac{\lenslot}{\Iup}.
\end{align}
In the following, only homogeneous traffic generation is considered, so we set
\begin{align}
\poissonrate = \begin{cases}
  \pgenup & \node{n} \neq \rootnode\\ 
  0 & \node{n} = \rootnode,
  \end{cases}
  , \forall 0 \leq i < \lensched.
\end{align}
but other traffic patterns can be obtained by choosing individual values for $\poissonrate$.

Besides the traffic generation itself, there is a probability $\Rx{n,i}$ of receiving a packet from a neighbor in a reception slot to be forwarded to the sink. This value is calculated from the probability $\Tx{\counterpart{n}(i),i}$ that the neighbor $\counterpart{n}(i)$ is transmitting in the given slot as%
\begin{align}%
  \Rx{n,i}\!=\!\begin{cases}
    \Tx{\counterpart{n}(i),i} & i \in \rxslots{n}\\
    0 & \text{else.}
  \end{cases}\label{math:forwarding}
\end{align}%
The forwarding is modeled by the Bernoulli part of the traffic model since at most one packet can be received per slot, so $\bernoulliprob = \Rx{n,i}$.

\subsection{Network Throughput}
The throughput of the network corresponds to the number of packets arriving at the sink $\rootnode$ per time. Since the sink does not generate traffic itself, the throughput can be calculated as
\begin{align}
  &\frac{\Atotal{0}}{\lenslot} = \lenslot^{-1} \cdot \sum_{i=0}^{\lensched-1}\Rx{0,i}.
\end{align}

\subsection{End-to-End Metrics}
The packet delivery ratio (PDR) $\Rup{n}$, that is the probability that a packet originating from node $\node{n}$ is finally received by the sink $\rootnode$, is given by the product of $\Paccept{n}$ over the path
\begin{align}
  \Rup{n} = \begin{cases}
  \Rup{\parent{n}} \cdot \Paccept{n} & \node{n} \neq \rootnode\\ 
  1 & \node{n} = \rootnode,
\end{cases}
\end{align}
where $\parent{n}$ is the parent of $\node{n}$ in the routing tree.

Similarly, the end-to-end delay $\Dup{n}$ is calculated as
\begin{align}
  \Dup{n} = \begin{cases}
  \Dup{\parent{n}} + \Exp{\Delay{n}} & \node{n} \neq \rootnode\\ 
  0 & \node{n} = \rootnode.
\end{cases}
\end{align}

\subsection{Comparing Model and Simulation}
\label{sect:simulation}
\begin{figure}[t]%
  \centering\includegraphics{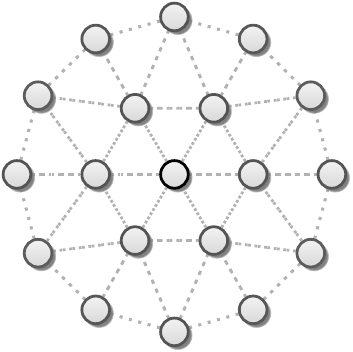}%
    \caption{A concentric network with 19 nodes.\label{fig:network}}
\end{figure} %

After evaluating the model for a single node only, we now build up a full multi-hop network to analyze its performance. An analytical model is very valuable for appreciating the validity of simulation results. Thus, we use the model to compare its results to the Contiki implementation of the Orchestra scheduling \cite{orchestra} for TSCH running in the COOJA simulator \cite{osterlind_cross-level_2006} with a maximum queue length of $K=16$. It is evaluated for a data-collection scenario in a concentric topology with 19 nodes as shown in Fig.~\ref{fig:network}. Sender-based Dedicated (SBD) Orchestra Slots are used, where every node has a dedicated transmission slot and thus, $\Disturb{i,\cup} = \emptyset$ and so the schedule is conflict-free and valid. Furthermore, the slotframe length is minimized to maximize the throughput under the given constraint.

\begin{figure}[tb]%
    \includegraphics{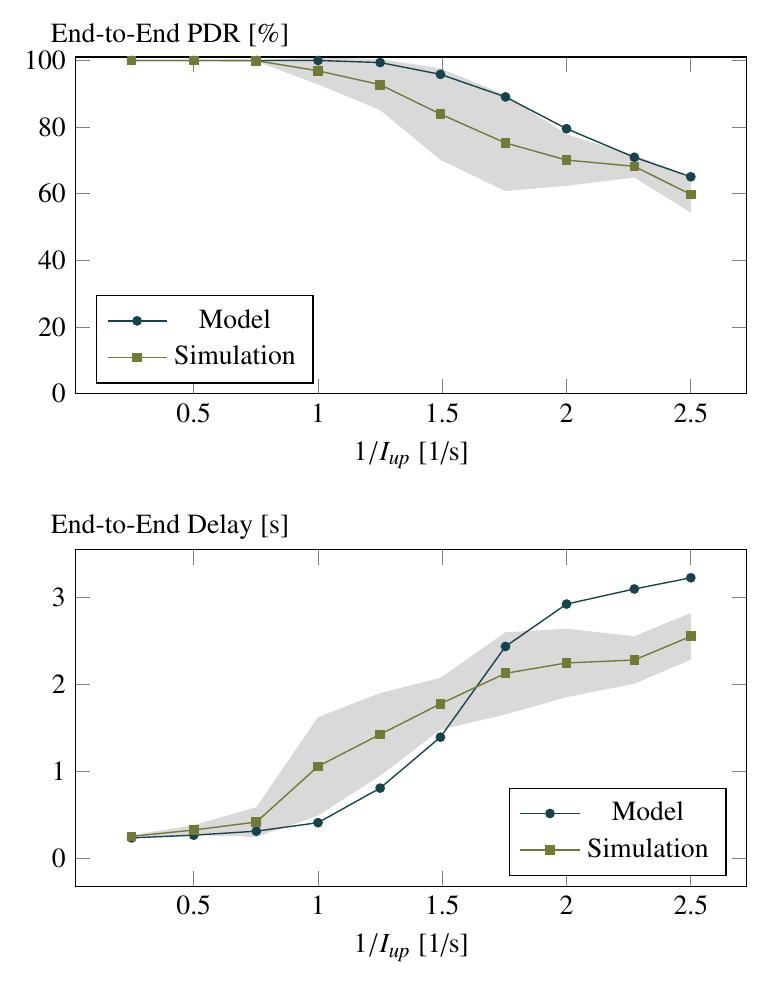}
    \caption{Comparing a Sender-based Dedicated Orchestra Slot Schedule in the Cooja Simulator and the model with $N=19$ and $K=16$. The average end-to-end packet delivery ratio and end-to-end delay are shown for the nodes in the outer circle.\label{fig:orchestra_model}}
\end{figure}

The results in Fig.~\ref{fig:orchestra_model} show the end-to-end packet delivery ratio (PDR) averaged over the nodes in the outer circle as well as the end-to-end delay. In the simulation, every node sends packets to the center with exponentially distributed intervals. After a warm-up phase of 15 minutes, 100 packets are monitored if they arrive and how long they take. The resulting mean and the 95\% confidence interval over 5 runs are shown in the plot.

While the simulation uses the Routing Protocol for Low power and Lossy Networks (RPL) \cite{rpl}, the analytical model uses a predefined tree routing as in \cite{modelarxiv}.
The PDR $\Rup{n}$ and the end-to-end delay $\Dup{n}$ are shown in the plot. Obviously, no confidence intervals are shown because no randomness is included in the calculation. The results show a good conformance between model and simulation and demonstrate the applicability of the analytical model for multi-hop networks as well as the validity of the Contiki simulation for the given scenario. The existing differences can be mainly traced back to the varying routing trees in the simulation runs.

\section{Comparing Multi-Hop Schedules}
\label{sect:schedules}
The analytical model is also very useful to assess new scheduling algorithms without the need to implement them for a full simulation stack. Orchestra is an appropriate schedule for many applications, especially since it requires no management traffic, but its main disadvantage is its inability to handle different traffic loads efficiently. In the given scenario, the inner nodes have to handle the aggregated traffic of all nodes in their sub tree, so they should be able to use more slots than the leafs.

In this section, two traffic-aware schedules are introduced for the data collection scenario presented in the previous section for a given tree. The main idea is that every node has enough slots for the traffic generated at that node as well as for forwarding the traffic of its children, assuming equal traffic distribution. More formally, if $\pdesc{n}$ denotes the number of proper descendants of node $\node{n}$, every node will get $\pdesc{n}+1$ transmission slots towards the sink. Algorithm~\ref{algo:pdesc} describes a distributed algorithm to calculate $\pdesc{n}$ for every node and also to make every node aware of the number of proper descendants of their children. For this, every node keeps the variables $\node{n}.\gamma$ for the number of proper descendants of this node $n$ and for each child $s$ its number of proper descendants $\node{n}.\gamma_s$. As before, $\rootnode$ denotes the sink node and $\parent{n}$ the index of the parent of node $\node{n}$.

The algorithm is basically a depth-first search. It is started by sending $\textproc{Forward}$ to $\rootnode$. After initializing $\node{0}.\gamma$, $\textproc{Forward}$ messages are sent until a leaf is reached. After this a \textproc{Backtrack} message is sent back to the parent. The parent increments its counters before continuing the depth-first search. After a subtree is fully handled, the number of nodes in this subtree is sent back to the parent in a \textproc{Backtrack} message.

In the presented distributed algorithms we assume all messages arrive correctly and in order. Otherwise, the algorithms would get a lot more complex. Associated problems and solutions for unreliable message exchange when building schedules are presented in \cite{formaldsme}.

\algdef{SE}[PROCEDURE]{OnMessage}{EndOnMessage}%
   [2]{\textbf{on message}\ \textproc{#1}\ifthenelse{\equal{#2}{}}{}{(#2)} \textbf{at \node{n} from \node{s}}}%
      {}%
\algtext*{EndOnMessage}%

\algdef{SE}[PROCEDURE]{Initialization}{EndInitialization}%
   {\textbf{initialization}}%
      {}%
\algtext*{EndInitialization}%

\newcommand{\send}[3]{\State \textbf{send} \textproc{#1}\ifthenelse{\equal{#2}{}}{}{(#2)} to #3}
\newcommand{\msgspace}{\vspace{0.3cm}}

\begin{algorithm}[tb]
    \caption{Determine proper descendants}
  \label{algo:pdesc}
  \begin{algorithmic}[1]
    \Initialization
      \State mark all children as unvisited $\forall\,\,\node{n}$
    \EndInitialization
\msgspace
    \OnMessage {Forward}{}
      \State $\node{n}.\gamma \leftarrow 0$
      \State \textproc{HandleNode}($\node{n}$)
    \EndOnMessage
\msgspace
    \OnMessage {Backtrack}{$j$}
      \State $\node{n}.\gamma_s \leftarrow j$ \Comment{Store proper descendants of children}
      \State $\node{n}.\gamma \leftarrow \node{n}.\gamma + j$
      \State \textproc{HandleNode}($\node{n}$)
    \EndOnMessage
\msgspace
      \Procedure {HandleNode}{$\node{n}$}
      \If{$\exists \text{ unvisited child } \node{u}$}
        \State mark $\node{u}$ as visited
        \send{Forward}{}{$\node{u}$}
      \ElsIf{$\node{n} \neq \rootnode$} \Comment{leaf or sub tree fully handled}
        \send{Backtrack}{$\node{n}.\gamma+1$}{$\node{\parent{n}}$}
      \EndIf
    \EndProcedure
  \end{algorithmic}
\end{algorithm}

The schedules presented in this section are meant to demonstrate the applicability of the analytical model and the advantage of traffic-aware schedules. However, they are not directly applicable to real-world applications due to their inability to cope with changes in the network topology and the traffic load without a complete recalculation. Yet, due to their simplicity, they serve as a starting point for more sophisticated traffic-aware schedules.

\subsection{Traffic-Aware Schedule (Single-Channel)}
The first traffic-aware scheduling algorithm as shown in Algorithm~\ref{algo:tasc} is based on the scheduling algorithm Type III from \cite{turau_tdma-schemes_2007}. As for Orchestra's Sender-based Dedicated Slots, only one node in the entire network is sending at every given point in time, so again $\Disturb{i,\cup} = \emptyset$. It is therefore denoted as single-channel scheduling, though, similar to Orchestra, channel hopping could be used to mitigate external interferences.

The algorithm is started by sending $\textproc{Track}(1)$ to $\rootnode$ after every node has performed the initialization. The algorithm is a depth-first search again, but instead of sending back the number of nodes in the subtree, multiple transmission slots are reserved towards the parent, one for handling the traffic of every proper descendant and one for the traffic generated at that node ($\pdesc{n}+1$). The transmission slot is recorded in the local $\txslots{n}$ and a message is sent to the parent $\node{\parent{n}}$ to record the reception slot in its $\rxslots{\parent{n}}$. When sending the \textproc{Track} message back to the parent, the number of already assigned slots is sent along as an offset for the next slot assignment.

The overall number of slots $\lensched$ in a slotframe for this schedule is calculated as
\begin{align}
  \lensched = 1 + \sum_{n = 1}^{N-1} \pdesc{n} + 1,
\end{align}
since every node (apart from $\rootnode$) performs $\pdesc{n}+1$ slot allocations.
The additional slot is the first slot of the slotframe that is usually left free for shared communication in most TSCH implementations, for example for management traffic. This also holds for the used Orchestra implementation.

\begin{algorithm}
    \caption{Traffic-Aware Schedule (Single-Channel)}
  \label{algo:tasc}

  \begin{algorithmic}[1]
    \Initialization
      \State $\txslots{n} \leftarrow () \,\,\forall\,\, \node{n}$
      \State $\rxslots{n} \leftarrow () \,\,\forall\,\, \node{n}$
      \State mark all children as unvisited $\forall\,\,\node{n}$
    \EndInitialization
\msgspace
    \OnMessage {Track}{$z$}
      \If{$\exists \text{ unvisited child } \node{u}$}
        \State mark $\node{u}$ as visited
        \send{Track}{$z$}{$\node{u}$}
      \ElsIf{$\node{n} \neq \rootnode$} \Comment{leaf or sub tree fully handled}
        \For{$i \leftarrow z,\ldots,z + \node{n}.\gamma$}
          \State $\txslots{n} \leftarrow \txslots{n} {^\frown} \left( i \right) $ \Comment{append to $\txslots{n}$}
          \State $\counterpart{n}(i) \leftarrow \parent{n}$
          \send{AssignRX}{$i$}{\node{\parent{n}}}
        \EndFor
        \send{Track}{$z+\node{n}.\gamma+1$}{$\node{\parent{n}}$}
      \EndIf
    \EndOnMessage
\msgspace
    \OnMessage {AssignRX}{$i$}
      \State $\rxslots{n} \leftarrow \rxslots{n} {^\frown} \left( i \right) $
      \State $\counterpart{n}(i) \leftarrow s$
    \EndOnMessage
  \end{algorithmic}
\end{algorithm}

\subsection{Traffic-Aware Schedule (Multi-Channel)}
In general, it is not required that only one node is sending at every point in time. Two pairs of nodes that are sufficiently apart, can send at the same time without interference. Secondly, nodes can communicate on different channels to avoid interference. This spatial and frequency diversity can be exploited to shorten the slotframe and therefore increase the throughput and lower the latency.

\begin{algorithm}[htb]
    \caption{Traffic-Aware Schedule (Multi-Channel)}
  \label{algo:tamc}
  \begin{algorithmic}[1]
    \Initialization
      \State $\txslots{n} \leftarrow () \,\,\forall\,\, \node{n}$
      \State $\rxslots{n} \leftarrow () \,\,\forall\,\, \node{n}$
      \State $\blockedchannels{n}(i) \leftarrow ()\,\, \forall\,\, 0 \leq i < \lensched, \,\,\forall\,\, \node{n}$
      \State mark all children as unvisited $\forall\,\,\node{n}$
    \EndInitialization
\msgspace
    \OnMessage {Track}{}
      \If{$\exists \text{ unvisited child } \node{u}$}
        \State mark $\node{u}$ as visited

        \State $\sigma \leftarrow \node{n}.\gamma_u + 1$\label{algoline:txslots}
        \For{$i \leftarrow 1,\ldots,\lensched-1$} \Comment{slot $0$ is reserved}
        \If{$i \notin \rxslots{n} \cup \txslots{n}$} \Comment{slot $i$ is idle}
            \State $\rxslots{n} \leftarrow \rxslots{n} {^\frown} \left( i \right) $ \Comment{append to $\rxslots{n}$}
            \State $\counterpart{n}(i) \leftarrow u$
            \State select $c \in \channels \setminus \blockedchannels{n}(i)$ \label{algoline:select}
            \State $\channel{\node{n}}(i) \leftarrow c$
            \send{AssignTX}{$i,c$}{$\node{u}$}
            \send{Block}{$i,c,\textit{true}$}{all $v \in \neighbors{n} \setminus \{\node{u}\}$}

            \State $\sigma \leftarrow \sigma-1$
            \If{$\sigma = 0$}
                \State \textbf{break} \Comment{all slots are assigned}\label{algoline:break}
            \EndIf
        \EndIf
        \EndFor

        \send{Track}{}{$\node{u}$}
      \ElsIf{$\node{n} \neq \rootnode$} \Comment{leaf or sub tree fully handled}
        \send{Track}{}{$\node{\parent{n}}$}
      \EndIf
    \EndOnMessage
\msgspace
    \OnMessage {AssignTX}{$i,c$}
      \State $\txslots{n} \leftarrow \txslots{n} {^\frown} \left( i \right) $
      \State $\counterpart{n}(i) \leftarrow s$
      \State $\channel{\node{n}}(i) \leftarrow c$
      \send{Block}{$i,c,\textit{true}$}{all $v \in \neighbors{n} \setminus \{\node{s}\}$}
    \EndOnMessage
\msgspace
      \OnMessage {Block}{$i,c,\text{forward}$}
      \State $\blockedchannels{n}(i) \leftarrow \blockedchannels{n}(i) \cup c$
      \If{forward}
        \send{Block}{$i,c,\textit{false}$}{$\parent{n}$}
      \EndIf
    \EndOnMessage
  \end{algorithmic}
\end{algorithm}

A corresponding distributed algorithm is given in Algorithm~\ref{algo:tamc}. In contrast to the previous algorithm, where the child determines the transmission slots towards the parent, in this algorithm the parent determines the reception slots in which it will expect the child to send. It is started by sending $\textproc{Track}$ to $\rootnode$. As in the previous algorithm, every node requires $\pdesc{n} + 1$ transmission slots. This is also calculated in line \ref{algoline:txslots} for a child $u$ when a node is first handled by its parent. Afterwards, $\sigma = \node{n}.\gamma_u + 1$ slot allocations are performed and the loop is always finished in line \ref{algoline:break} if we assume there are always enough channels (see below, otherwise line \ref{algoline:select} would fail) and $\lensched$ is chosen as follows.
For every allocation, a previously unused time slot is searched, recorded at sender and receiver and the used channel is blocked in the 2-hop neighborhood. For this, every node maintains a set of blocked channels for every slot $\blockedchannels{n}(i)$. This search will always be successful if we choose the slotframe length as
\begin{align}
  \lensched = 1 + \max_{n\,=\,0,\ldots,N-1}\begin{cases}
      2\cdot\pdesc{n} + 1 & \node{n} \neq \rootnode\\
      \pdesc{n} & \node{n} = \rootnode
  \end{cases},
\end{align}
because the root requires that the slotframe has at least one slot for receiving (potentially forwarded) traffic from every proper descendant. For all other nodes this holds, too, but in addition the same number of slots is required for forwarding and one slot for transmitting the traffic generated at that node.
Again, the additional slot is the shared first slot in the slotframe. 
Since we assume enough channels are available to avoid conflicts, the slotframe length corresponds to the requirement of the node with the largest required slotframe length.

In contrast to the other algorithms, transmissions can take place simultaneously without or with very little interference by assigning a dedicated channel or a non-conflicting channel hopping sequence to every conflicting link in a neighborhood. 
In the presented algorithm, this is ensured by signaling every slot assignment to all neighbors $\neighbors{n}$ of the sender and the receiver as well as the parents of the neighbors. Thereby, the four interference constellations in Fig.~\ref{fig:rel} are avoided. This idea is similar to the slot allocation handshake of DSME as well as the DeBraS scheduling algorithm \cite{municio_decentralized_2016} and is especially important for dense networks.

\begin{figure}[tb]%
  \centering\includegraphics{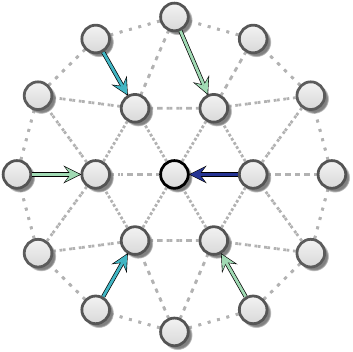}%
    \caption{A possible coloring for a set of conflicting links.\label{fig:coloring}}
\end{figure} %
  
\begin{figure}[tb]%
  \centering\includegraphics{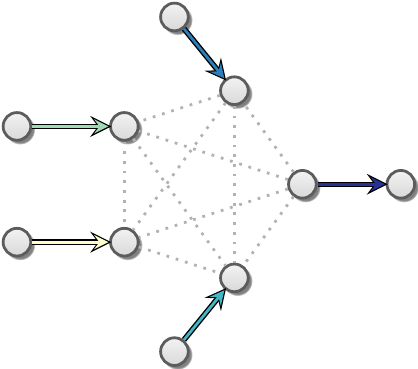}%
    \caption{A schedule that requires a coloring with five colors.\label{fig:coloringfive}}
\end{figure}

In general, this is a graph coloring problem where $\Li$ is the set of vertices and $\Disturb{i,\cup}$ is the set of edges. Executing the presented heuristic algorithm for the first slot results in the coloring shown in Fig.~\ref{fig:coloring}. Here, only three channels are required, so the 16 channels available for IEEE~802.15.4 in the $2.4\,\text{GHz}$ band are more than sufficient. In general, however, no upper bound can be given for the number of required channels, because the conflict graph is not necessarily planar. Fig.~\ref{fig:coloringfive} shows an example that requires five channels and could be extended to an arbitrary number of channels. While in general finding a valid coloring with at most 16 channels is not ensured, it is usually more than sufficient in real-world applications. 

\subsection{Evaluation}

\begin{figure*}[bt]%
  \centering\includegraphics{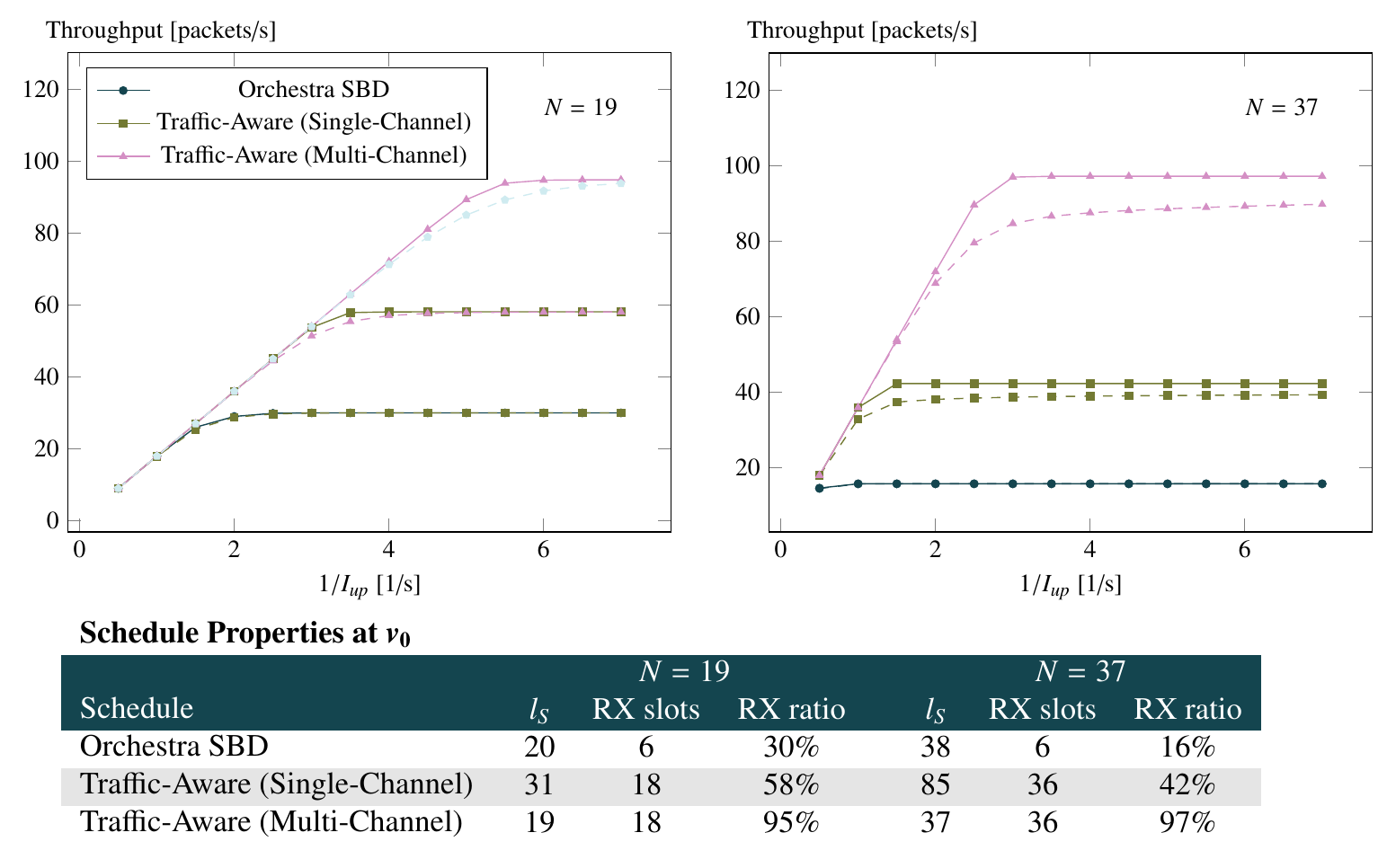}%
    \caption{Throughput for the Sender-based Dedicated Orchestra Schedule and the two presented traffic-aware schedules over the packet sending interval for a network of $N=19$ and $N=37$ nodes aligned in concentric circles. The solid lines are for $K=16$, the dashed for $K=6$.\label{fig:throughput}. In addition, the table presents schedule properties at the root node $\rootnode$.}
\end{figure*} %

In Fig.~\ref{fig:throughput}, the achievable throughput for different schedules is compared for multiple scenarios, i.e. for a network of $N=19$ and $N=37$ nodes in a concentric topology and for maximum queue sizes of $K=6$ (dashed) and $K=16$ (solid). For this, the number of packets received by the sink $\rootnode$ is plotted over the packet generation rate of every node. For low rates, the network is able to handle the complete traffic, so the throughput rises linearly. For high rates, the networks is saturated and increasing the rate does not increase the throughput anymore.

In general, the traffic-aware schedules provide significantly more throughput than Orchestra, while the multi-channel schedule has an even higher throughput than the single-channel schedule. The higher throughput can be explained by considering that nodes higher in the tree that need to handle more traffic have more slots compared to nodes with low traffic demand.

It is also apparent in Fig.~\ref{fig:throughput} that the maximum queue length $K$ is most significant in the transition section, in conformance to Sect.~\ref{sect:queuedist}. Compared to the applied schedule, it has a lower impact on the throughput, but the difference is larger for $N=37$ than for $N=19$ due to the higher number of hops.

Also, when comparing different network sizes, we see that the saturation is reached for smaller rates. This is expected, because the same packet generation rate applied at more nodes leads to more overall traffic. Secondly, the single-channel schedule and even more Orchestra SBD show a significantly lower throughput, because having more nodes requires a higher number of slots during which the sink is idle. This does not hold for the multi-channel schedule, because spatial reuse is possible.

In the table of Fig.~\ref{fig:throughput}, the slotframe lengths are given together with the number and ratio of reception slots at the sink $\rootnode$. For Orchestra SBD, every neighbor of $\rootnode$ has only one slot, so $\rootnode$ has a long phase of inactivity. For the traffic-aware schedules, the inner nodes get more slots because of their higher traffic load. Thus, the sink can receive more packets per slotframe resulting in a higher throughput. The missing slot in the last row is due to the shared slot. In the multi-channel schedule, the sink can even potentially receive data traffic in every slot, apart from the first one. This comparison again explains the difference in throughput for the different network sizes, because comparative to the network with 19 nodes, the RX ratio is significantly lower for $N=37$ and the Orchestra SBD and single-channel schedules, while it is even larger for the multi-channel schedule where the shared slot has a lower impact due to the longer slotframe length.

\section{Possible Extensions}
\label{sect:extensions}
\subsection{IEEE 802.15.4 DSME}
While the focus of this paper is TSCH due to its higher flexibility and broader usage, the model can easily applied to DSME, too. The main difference between TSCH and DSME is the availability of distributed management procedures for setting up schedules in the latter, but since the presented model analyzes schedules in a steady state, the only major difference is the slot structure.
In TSCH, the time slots can be arbitrarily dedicated as contention-free and contention-access slots. DSME has a less flexible structure. It consists of one beacon slot, 8 contention-access slots and 7 contention-free slots aligned in a fixed, yet configurable, repeated sequence. The contention-access phase is usually used for management, similar to, but longer than the extra slot used in this paper.
Secondly, the slots are usually shorter, because TSCH requires some extra time in every slot due to its time synchronization procedure, while DSME uses a dedicated beacon slot (see \cite{juc_energy_2016} for details). Therefore, the timing of the model has to be adapted and the schedule has to account for the slots not used for contention free communication.

Furthermore, due to the less flexible structure, a scheduling algorithm such as Algorithm \ref{algo:tamc} would lead to excess slots that are not utilized. Therefore, a more elaborate algorithm is required to achieve optimal performance by evenly distributing the excess slots.

\subsection{Other Sources of Packet Loss}
\label{sect:otherloss}
As outlined in Sect.~\ref{sect:requirements}, most packet losses in TSCH and DSME networks are queue drops. However, in practice other sources of packet loss are inevitable. The consideration of packet loss during the transmission can be integrated in the presented model by replacing equation (\ref{math:forwarding}) with%
\begin{align}%
  \Rx{n,i}\!=\!\begin{cases}
    \Tx{\counterpart{n}\!(i),i}\!\left(1\!-\!\PER{b,(\counterpart{n}\!(i),n)}\!(t)\right)\!\!\!\!\!\!& i \in \rxslots{n}\\
    0 & \text{else,}
  \end{cases}
\end{align}%
where $\PER{b,(\counterpart{n}(i),n)}(t)$ is the probability that a packet transmission with $b$ byte fails at time $t$.

An example for a time-independent calculation of $\PER{b,(m,n)}$ can be found in \cite{modelarxiv} and in \cite{di_marco_effects_2013} a model for Rayleigh-lognormal fading in IEEE 802.15.4 networks is given. Time dependence is required for accurate modeling of external interferences or fading channels, but that is out of the scope of this paper.

Furthermore, since the model assumes no losses on the physical layer, retransmissions are not considered. If other sources of packet loss are considered, retransmissions have to be integrated in the model by adding transitions in the Markov chain for maintaining the queue level after a failed transmission. For the considered scenario, these transitions would never be taken and thus not change the results.

\section{Conclusion}
\label{sect:conclusion}

The paper presents an analytical approach for the assessment of wireless mesh networks that use a collision-free TDMA. 
A queuing model based on a Markov chain is proposed that models forwarding traffic and irregular slot schedules accurately, in contrast to the well-known M/D/1/K model.
These models are linked together to build up a multi-hop model of the whole network for calculating packet delivery ratio, end-to-end delay and throughput.

The results demonstrate the increased accuracy compared to the M/D/1/K model and illustrate the effect of a finite queue by showing the queue level distribution. For evaluating the multi-hop model, a data-collection scenario is applied. The analytical model is compared to a simulation of the Orchestra schedule, showing good conformance. 
Finally, two distributed traffic-aware scheduling algorithms are presented. The higher throughput achieved by traffic-awareness is demonstrated and the influence of the maximum queue length is shown.

The calculations were conducted by means of the Portable, Extensible Toolkit for Scientific Computation (PETSc) \cite{petsc-efficient,petsc-user-ref}, the simulations with COOJA \cite{osterlind_cross-level_2006} and SimPy \cite{simpy}. The open source implementation of the models can be accessed at \cite{modelimpl}. 

Overall, the proposed analytical model, together with its software implementation, is a useful tool for testing new ideas while developing new slot schedules. It is also very helpful for practitioners who want to estimate the performance of wireless mesh networks.

\section*{Appendix A: Expected Value of $\Arr{n,i}$}
  \label{app:qmean}
\noindent The expected value of $\Arr{n,i}$ is calculated as
\begin{align}%
  \begin{aligned}
    &\Exp{\Arr{n,i}} = \sum_{k=0}^\infty k\cdot \Prob{\Arr{n,i} = k}\\
      &\begin{aligned}
        = \sum_{k=1}^\infty k \left(\left(1\!-\!\bernoulliprob\right)\cdot\Prob{\Poi = k}\right.\\
        \left.\hspace{0.3cm}+ \,\bernoulliprob\cdot\Prob{\Poi = k-1}\right)
    \end{aligned}\\
  &= \left(1-\bernoulliprob\right) \cdot \left(\sum_{k=1}^\infty k \cdot \Prob{\Poi = k}\right) \\
    & \quad + \bernoulliprob\cdot\sum_{k=1}^\infty k \cdot \Prob{\Poi = k-1}
  \end{aligned}
\end{align}
so with
\begin{align}%
  \begin{aligned}
  \!\sum_{k=1}^\infty k\!\cdot\!\Prob{\Poi = k} &\!= \sum_{k=1}^\infty k\cdot \frac{\poissonrate^{k}}{k!} e^{-\poissonrate}\\
  &\!= \poissonrate\!\cdot\! e^{-\poissonrate} \sum_{k=1}^\infty \frac{\poissonrate^{k-1}}{(k-1)!} \\
  &\!= \poissonrate\!\cdot\! e^{-\poissonrate} \sum_{j=0}^\infty \frac{\poissonrate^{j}}{j!}\\
  &\!= \poissonrate\!\cdot\! e^{-\poissonrate} e^{\poissonrate} = \poissonrate
  \end{aligned}
\end{align}
and
\begin{align}%
  \nonumber &\sum_{k=1}^\infty k \cdot \Prob{\Poi = k-1} = \sum_{k=1}^\infty k\cdot \frac{\poissonrate^{k-1}}{(k-1)!} e^{-\poissonrate}\\
  \nonumber &= e^{-\poissonrate} \!\left(\!\!\left(\sum_{k=1}^\infty \frac{\poissonrate^{k-1}}{(k-1)!}\right)\!+\!\sum_{k=1}^\infty(k-1)\frac{\poissonrate^{k-1}}{(k-1)!}\right)\\
  &= e^{-\poissonrate} \!\left(\!\!\left(\sum_{k=1}^\infty \frac{\poissonrate^{k-1}}{(k-1)!}\right)\!+\!\poissonrate \sum_{k=2}^\infty\frac{\poissonrate^{k-2}}{(k-2)!}\right)\\
  \nonumber &= e^{-\poissonrate} \!\left(\!\!\left(\sum_{j=0}^\infty \frac{\poissonrate^{j}}{j!}\right) + \poissonrate \sum_{j=0}^\infty\frac{\poissonrate^{j}}{j!}\right)\\
    \nonumber &= e^{-\poissonrate} \left(e^{\poissonrate} + \poissonrate\cdot e^{\poissonrate}\right) = 1 + \poissonrate
\end{align}
we finally get
\begin{align}
  &\Exp{\Arr{n,i}} = \left(1\!-\!\bernoulliprob\right) \poissonrate\!+\!\bernoulliprob \left(\poissonrate\!+\!1\right).
\end{align}%

\section*{Appendix B: Stationary Distribution}
\label{sect:steady}

The stationary distribution of the presented Markov chain for node $\node{n}$ is denoted as
\begin{align}
  \cvect = \left(\csteady{n,\left\lfloor\tfrac{j}{\lensched}\right\rfloor,j\%\lensched}\right)_{j=0,\ldots,(K+1)\cdot\lensched-1}
\end{align}
where all $0 \leq \csteady{n,q,i} \leq 1$ and $\cvect \cdot e = 1$ with
\begin{align}
e = \left(\begin{array}{cccc}1 & 1 & \ldots & 1\end{array}\right)^T,
\end{align}
that is the normalization criterion that the probabilities have to sum up to $1$.
The stationary distribution is calculated as the solution of $\cvect P = \cvect$, where $P$ is the transition probability matrix
\begin{align}
  \begin{aligned}
    P = \left[p_{j,k}\right]_{((K+1)\cdot\lensched) \times ((K+1)\cdot\lensched)}\\
    p_{j,k} = \qqtr{\left\lfloor\tfrac{j}{\lensched}\right\rfloor,j\%\lensched}{\left\lfloor\tfrac{k}{\lensched}\right\rfloor,k\%\lensched}.
  \end{aligned}
\end{align}
This can be rewritten as
\begin{align}
  \cvect(I - P) = 0 \Leftrightarrow (I - P)^T \cvect^T = 0,
\end{align}
with the identity matrix I. This is a homogeneous system 
of $(K+1)\cdot\lensched$ linear equations and the same number of unknowns. 

\subsection*{Irreducibility}
\begin{figure}[h!]%
\centering\includegraphics{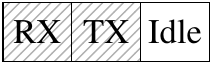}%
    \caption{A schedule with unreachable states $\qstate{1,2}$.\label{fig:reducible}}
\end{figure}

Consider the example in Fig.~\ref{fig:reducible} with $\lensched=3$, $K = 1$, $\pgenup = 0$ and $\Rx{n,0} > 0$. Since nothing is generated and every packet received in slot $0$ is immediately sent out again, the state $\qstate{1,2}$ is never visited when starting with $q=0$.
It is also possible to construct more complex examples where blocks of states exist that are linked together, but are not reachable from any state with $q=0$. This property makes the Markov chain reducible and thus a unique $\cvect$ is not guaranteed by the following proof. Finding these states corresponds to finding the states that are not in the strongly connected set that contains $\qstate{0,0}$.

Since it is meaningless for the application to assign any $\csteady{n,q,i} > 0$ to those states, they are set to zero and the corresponding columns and rows are removed from $P$, making the Markov chain and $P$ irreducible, that is there is no permutation of the rows and columns of $P$ resulting in
\begin{align}
  \left(\begin{array}{cc}
A & B\\
0 & D
\end{array}\right),
\end{align}
with the square matrices $A$ and $D$ and the matrix $0$ with all elements zero.

\subsection*{Rank of $I-P$}
In the following we prove that the matrix $Q = I - P$ has rank $n-1$ if $P$ is irreducible. The proof goes along the lines of \cite{chakravarti_characterization_1975}. Since the outgoing transitions of a state have to sum up to one, that is 
\begin{align}
\forall \qstate{q_1,i_1}\in\,\states:\!\!\!\!\sum_{\qstate{q_2,i_2}\,\in\,\states}\!\!\!\!\!\! \qqtr{q_1,i_1}{q_2,i_2} = 1,
\end{align}
it holds that $P e = 1$, thus $(I - P) e = 0$, so the matrix $Q$ has at least one zero eigenvalue, is therefore singular and its rank is at most $n-1$.

Assuming a rank of $n-2$ or less, then there is at least one vector $x$ with $Q x = 0$ that is orthogonal to $e$, i.e. $x^T e = 0$. Thus, all linear combinations of $x$ and $e$ are also in the null space of $Q$. 
In particular this holds for $d = x - m\cdot e$ where $m$ is the minimum entry of $x$. Then at least one, but not all, elements of $d$ are $0$ and all others are larger than $0$. Note that this would not hold for $x$ parallel to $e$. However, this is not possible since $x^T e = 0$.

Since permutations do not change the rank, we assume without loss of generality the first $u>0$ elements of $d$ are positive and the remaining $n-u$ elements are zero. It holds%
\begin{multline}
Q d = 0  \Leftrightarrow (I-P) d = Id - Pd = 0\\ \Leftrightarrow P d = I d \Leftrightarrow Pd = d.
\end{multline}
This can be partitioned as
\begin{align}
  P d = \left(\begin{array}{cc}
A & B\\
C & D
    \end{array}\right)
    \left(\begin{array}{c}d_1\\\vdots\\d_u\\0\\\vdots\\0\end{array}\right) = 
    \left(\begin{array}{c}d_1\\\vdots\\d_u\\0\\\vdots\\0\end{array}\right),
\end{align}
with $C$ of dimension $(n-u) \times u$. Therefore, it holds
\begin{align}
C \left(\begin{array}{c}d_1\\\vdots\\d_u\end{array}\right) = \left(\begin{array}{c}0\\\vdots\\0\end{array}\right).
\end{align}
For $d_i > 0\,\,\,\forall 1 \leq i \leq u$ this is only possible if all entries of $C$ are zero or there are negative entries in $C$. The first one contradicts the irreducibility of $P$ and the second one would require the existence of negative probabilities in $P$. Therefore, $I-P$ has rank $n-1$. Since the rank is maintained by transposition, the matrix $(I-P)^T$ has rank $n-1$, too.

\subsection*{Solution of $cP = c$}
Since $(I-P)^T$ has rank $n-1$, the homogeneous system of linear equations $(I-P)^T c^T = 0$ has a solution space of dimension one, so since $c \cdot e = 1$, there is a unique stationary distribution. Furthermore, if we find any vector $y \neq 0$ with $(I-P)^T y^T = 0$, we can get $c$ by normalization.

Though most methods for the numerical solution of systems of linear equations are tailored to the handling of regular matrices, many can be adapted for the singular case as presented in \cite{stewart_introduction_1994}. For our application, the implementation of the generalized minimal residual method (GMRES) in PETSc \cite{petsc-user-ref} with SOR preconditioning and initial guess $0.5 e$ turned out to work very well. For more background about why and when GMRES is applicable to Markov chains see Section 4.4.4 in \cite{stewart_introduction_1994}.


\begin{backmatter}

\printnoidxglossary[type=symbols]

\section*{Abbreviations}
CAP: Contention Access Period;
CSMA/CA: Carrier Sense Multiple Access/Collision Avoidance;
DSME: Deterministic and Synchronous Multi Channel Extension;
GMRES: Generalized Minimal Residual Method;
PDR: Packet Delivery Ratio;
PER: Packet Error Ratio;
PETSc: Portable, Extensible Toolkit for Scientific Computation;
RPL: Routing Protocol for Low Power and Lossy Networks;
RX: Reception;
SBD: Sender-Based Dedicated;
TDMA: Time-Division Multiple Access;
TSCH: Time Slotted Channel Hopping;
TX: Transmission

\section*{Availability of software}
  The software implementing the presented model is available at https://github.com/koalo/AnalyticalMultiHop under the terms of the GNU General Public License v3.0. It is written in C++ and based on the Portable, Extensible Toolkit for Scientific Computation (PETSc) available for Linux and Windows.

\section*{Competing interests}
The authors declare that they have no competing interests.

\section*{Funding}
Funded by the Deutsche Forschungsgemeinschaft (DFG, German Research Foundation) - Projektnummer 392323616 and the Hamburg University of Technology (TUHH) in the funding programme Open Access Publishing. 

\section*{Author's contributions}
FK developed and evaluated the presented model and wrote the draft as main author. VT provided many useful proposals and significantly contributed to the revision of the whole paper. All authors read and approved the final manuscript.


\bibliographystyle{aps_eurasip} 
\bibliography{tdma_model}      

\begin{thebibliography}{48}
\providecommand{\enquote}[1]{``#1''}
\providecommand{\url}[1]{{\tt #1}}
\providecommand{\href}[2]{#2}


\bibitem[1]{odonovan_ginseng_2013}
T O'donovan, J Brown, F Büsching, A Cardoso, J Cecílio, J D Ó, P Furtado, P Gil, A Jugel, W-B Pöttner, \textit{et al.}, The {GINSENG} {System} for {Wireless} {Monitoring} and {Control}: {Design} and {Deployment} {Experiences}. ACM Trans. Sen. Netw. 10(1), 4:1--4:40 (2013).

\bibitem[2]{pfahl_holistic_2014}
A Pfahl, M Randt, F Meier, M Zaschke, C P W Geurts, M Buselmeier, {A Holistic Approach for Low Cost Heliostat Fields}, in Proceedings of the 20th {International} {Conference} on {Concentrated} {Solar} {Power} and {Chemical} {Energy} {Technologies} {(SolarPACES)}, Beijing, China, Sep 2014.

\bibitem[3]{meiermodel}
F Meier, V Turau, {An Analytical Model for Fast and Verifiable Assessment of Large Scale Wireless Mesh Networks}, in {Proceedings of the 11th International Conference on the Design of Reliable Communication Networks (DRCN)}, Kansas City, MO, USA, Mar 2015.

\bibitem[4]{802154}
IEEE, {IEEE 802.15.4\texttrademark-2015 - IEEE Standard for Local and metropolitan area networks--Part 15.4: Low-Rate Wireless Personal Area Networks (WPANs)}, 2016.

\bibitem[5]{towards_openDSME}
M K\"ostler, F Kauer, T L\"ubkert, V Turau, {Towards an Open Source Implementation of the IEEE 802.15.4 DSME Link Layer}, in Proceedings of the 15. GI/ITG KuVS Fachgespr\"ach Sensornetze, ed. by J. Scholz and A. von Bodisco{}, Augsburg, Germany, Sep 2016.

\bibitem[6]{orchestra}
S Duquennoy, B Al Nahas, O Landsiedel, T Watteyne, Orchestra: {Robust} {Mesh} {Networks} {Through} {Autonomously} {Scheduled} {TSCH}, in Proceedings of the 13th {ACM} {Conference} on {Embedded} {Networked} {Sensor} {Systems}, Seoul, South Korea, Nov 2015.

\bibitem[7]{modelimpl}
F Kauer, {Analytical Model for IEEE 802.15.4 Mesh Networks - Implementation} (2017), GitHub Repository,  https://github.com/koalo/AnalyticalMultiHop

\bibitem[8]{pathak_survey_2011}
P Pathak, R Dutta, A {Survey} of {Network} {Design} {Problems} and {Joint} {Design} {Approaches} in {Wireless} {Mesh} {Networks}. IEEE Communications Surveys \& Tutorials 13(3), (2011).

\bibitem[9]{zheng_comprehensive_2006}
J Zheng, M J Lee, {A Comprehensive Performance Study of IEEE 802.15.4}, in {Sensor Network Operations}, ed. by S. Phoha, T. LaPorta, and C. Griffin{}, {John Wiley \& Sons}, Hoboken, NJ, USA, pp.~218--236, 2005.

\bibitem[10]{dutta_trio_2006}
P Dutta, J Hui, J Jeong, S Kim, C Sharp, J Taneja, G Tolle, K Whitehouse, D Culler, {Trio: Enabling Sustainable and Scalable Outdoor Wireless Sensor Network Deployments}, in {Proceedings of the 5th International Conference on Information Processing in Sensor Networks (IPSN)}, Nashville, TN, USA, Apr 2006.

\bibitem[11]{misic_performance_2006}
J Mi\v{s}i\'{c}, S Shafi, V B Mi\v{s}i\'{c}, {Performance of a Beacon Enabled IEEE 802.15.4 Cluster with Downlink and Uplink Traffic}. {IEEE Transactions on Parallel and Distributed Systems} {17}(4), (2006).

\bibitem[12]{dimarco_analytical_2012}
P Di Marco, P Park, C Fischione, K H Johansson, {Analytical Modeling of Multi-hop IEEE 802.15.4 Networks}. {IEEE Transactions on Vehicular Technology} {61}(7), (2012).

\bibitem[13]{alderisi_simulative_2015}
G Alderisi, G Patti, O Mirabella, L L Bello, {Simulative Assessments of the {IEEE} 802.15.4e {DSME} and {TSCH} in Realistic Process Automation Scenarios}, in Proceedings of the 13th {International} {Conference} on Industrial {Informatics} ({INDIN}), Cambridge, UK, Jul 2015.

\bibitem[14]{guglielmo_performance_2014}
D D Guglielmo, A Seghetti, G Anastasi, M Conti, {A Performance Analysis of the Network Formation Process in {IEEE} 802.15.4e {TSCH} Wireless Sensor/Actuator Networks}, in Proceedings of the 19th {IEEE} {Symposium} on {Computers} and {Communications} ({ISCC}), Madeira, Portugal, Jun 2014.

\bibitem[15]{sahoo_novel_2017}
P Sahoo, S Pattanaik, S-L Wu, A {Novel} {IEEE} 802.15.4e {DSME} {MAC} for {Wireless} {Sensor} {Networks}. Sensors 17(1), (2017).

\bibitem[16]{jeong_performance_2012}
W-C Jeong, J Lee, {Performance Evaluation of {IEEE} 802.15.4e {DSME} {MAC} Protocol for Wireless Sensor Networks}, in Proceedings of the {First} {IEEE} {Workshop} on {Enabling} {Technologies} for {Smartphone} and {Internet} of {Things} ({ETSIoT}), Seoul, South Korea, Jun 2012.

\bibitem[17]{gronkvist_throughput_2004}
J Gronkvist, J Nilsson, D Yuan, {Throughput of Optimal Spatial Reuse {TDMA} for Wireless Ad-hoc Networks}, in Proceedings of the 59th {IEEE} Vehicular {Technology} {Conference} ({VTC}), Milan, Italy, May 2004.

\bibitem[18]{bjorklund_resource_2003}
P Bjorklund, P Varbrand, {Di Yuan}, Resource {Optimization} of {Spatial} {TDMA} in {Ad} {Hoc} {Radio} {Networks}: {A} {Column} {Generation} {Approach}, in Proceedings of the 22th {Annual} {Joint} {Conference} of the {IEEE} {Computer} and {Communications} {Societies} ({INFOCOM}), San Francisco, CA, USA, Mar 2003.

\bibitem[19]{ng_queueing_2008}
C-H Ng, S Boon-Hee, Queueing {Modelling} {Fundamentals}: {With} {Applications} in {Communication} {Networks}, 2nd edn. ({John Wiley \& Sons}, Hoboken, NJ, USA, 2008).

\bibitem[20]{seo_explicit_2014}
D-W Seo, Explicit {Formulae} for {Characteristics} of {Finite}-{Capacity} {M}/{D}/1 {Queues}. ETRI Journal 36(4), 609--616 (2014).

\bibitem[21]{macgregor_smith_properties_2011}
J MacGregor Smith, {Properties and Performance Modelling of Finite Buffer {M}/{G}/1/{K} Networks}. Computers \& Operations Research 38(4), 740--754 (2011).

\bibitem[22]{khan_delay_1998}
K Khan, H Peyravi, {Delay and Queue Size Analysis of {TDMA} with General Traffic}, in Proceedings of the 6th International Symposium on Modeling, Analysis and Simulation of Computer and Telecommunication Systems (MASCOTS), Montreal, Canada, Jul 1998.

\bibitem[23]{dunkels_contiki_2004}
A Dunkels, B Gr\"onvall, T Voigt, Contiki - {A} {Lightweight} and {Flexible} {Operating} {System} for {Tiny} {Networked} {Sensors}, in Proceedings of the 29th International Conference on Local Computer Networks, Tampla, FL, USA, Nov 2004.

\bibitem[24]{dujovne_6tisch_2017_sf0}
D Dujovne, L Grieco, M Palattella, N Accettura, 6TiSCH 6top {Scheduling} {Function} {Zero} ({SF}0), Internet Engineering Task Force,   (2017).

\bibitem[25]{anamalamudi_scheduling_2017}
S Anamalamudi, M Zhang, A Sangi, C Perkins, S V R Anand, Scheduling {Function} {One} ({SF}1) for hop-by-hop {Scheduling} in 6tisch {Networks}, Internet Engineering Task Force (IETF),   (2017).

\bibitem[26]{wirelesshart}
IEC, {IEC 62591:2010 - Industrial communication networks - Wireless communication network and communication profiles - WirelessHART\texttrademark}, 2010.

\bibitem[27]{pottner_constructing_2014}
W-B Pöttner, H Seidel, J Brown, U Roedig, L Wolf, Constructing {Schedules} for {Time}-{Critical} {Data} {Delivery} in {Wireless} {Sensor} {Networks}. ACM Trans. Sen. Netw. 10(3), 44:1--44:31 (2014).

\bibitem[28]{palattella_optimal_2013}
M R Palattella, N Accettura, L A Grieco, G Boggia, M Dohler, T Engel, On {Optimal} {Scheduling} in {Duty}-{Cycled} {Industrial} {IoT} {Applications} {Using} {IEEE}802.15.4e {TSCH}. IEEE Sensors Journal 13(10), 3655--3666 (2013).

\bibitem[29]{accettura_decentralized_2015}
N Accettura, E Vogli, M R Palattella, L A Grieco, G Boggia, M Dohler, Decentralized {Traffic} {Aware} {Scheduling} in 6TiSCH {Networks}: {Design} and {Experimental} {Evaluation}. IEEE Internet of Things Journal 2(6), 455--470 (2015).

\bibitem[30]{hwang_distributed_2017}
R-H Hwang, C-C Wang, W-B Wang, A {Distributed} {Scheduling} {Algorithm} for {IEEE} 802.15.4e {Wireless} {Sensor} {Networks}. Comput. Stand. Interfaces 52(C), 63--70 (2017).

\bibitem[31]{soua_wave_2016}
R Soua, P Minet, E Livolant, {Wave: a Distributed Scheduling Algorithm for Convergecast in {IEEE} 802.15.4e {TSCH} Networks}. Transactions on Emerging Telecommunications Technologies 27(4), 557--575 (2016).

\bibitem[32]{soua_wave_2016_ext}
R Soua, P Minet, E Livolant, {Wave: a Distributed Scheduling Algorithm for Convergecast in {IEEE} 802.15.4e {TSCH} Networks (Extended Version)},  Research Report RR-8661, Inria,  (2015).

\bibitem[33]{municio_decentralized_2016}
E Municio, S Latré, {Decentralized Broadcast-based Scheduling for Dense Multi-hop {TSCH} Networks}, in Proceedings of the Workshop on Mobility in the Evolving Internet Architecture (MobiArch), New York, USA, Oct 2016.

\bibitem[34]{ojo_energy_2017}
M Ojo, S Giordano, G Portaluri, D Adami, M Pagano, {An Energy Efficient Centralized Scheduling Scheme in {TSCH} Networks}, in Proceedings of the {IEEE} {International} {Conference} on {Communications} {Workshops} ({ICC} {Workshops}), Paris, France, May 2017.

\bibitem[35]{formaldsme}
F Kauer, M K\"ostler, T L\"ubkert, V Turau, {Formal Analysis and Verification of the IEEE 802.15.4 DSME Slot Allocation}, in Proceedings of the 19th ACM International Conference on Modeling, Analysis and Simulation of Wireless and Mobile Systems (MSWIM), Malta, Nov 2016.

\bibitem[36]{Telematik_Adhoc-Now_2015}
G Siegemund, V Turau, C Weyer, A Dynamic Topology Control Algorithm for Wireless Sensor Networks, in Proceedings of the International Conference on Ad-hoc, Mobile and Wireless Networks (ADHOC-NOW), Athens, Greece, Jun 2015.

\bibitem[37]{wirelessnerc}
T Kuruganti, W Dykas, W Manges, T Flowers, M Hadley, P Ewing, T King, {Wireless System Considerations When Implementing NERC Critical Infrastructure Protection Standards},  Oak Ridge National Laboratory, TN, USA, Flowers Control Center Solutions, Todd Mission, TX, USA, Pacific Northwest National Laboratory,  Richland, WA, USA (2009).

\bibitem[38]{simpy}
{Team SimPy}, {SimPy - Discrete Event Simulation for Python} (2017),  https://simpy.readthedocs.io, Accessed 14 Jul 2017.

\bibitem[39]{osterlind_cross-level_2006}
F Osterlind, A Dunkels, J Eriksson, N Finne, T Voigt, Cross-{Level} {Sensor} {Network} {Simulation} with {COOJA}, in Proceedings of the 31st International Conference on Local Computer Networks, Tampla, FL, USA, Nov 2006.

\bibitem[40]{rpl}
R Alexander, A Brandt, J Vasseur, J Hui, K Pister, P Thubert, P Levis, R Struik, R Kelsey, T Winter, {RPL: IPv6 Routing Protocol for Low-Power and Lossy Networks}, Internet Engineering Task Force,   (2012).

\bibitem[41]{modelarxiv}
F Meier, V Turau, {Analytical Model for IEEE 802.15.4 Multi-Hop Networks with Improved Handling of Acknowledgements and Retransmissions} (2015), arXiv:1501.07594 [cs.NI],  http://arxiv.org/abs/1501.07594

\bibitem[42]{turau_tdma-schemes_2007}
V Turau, C Weyer, {TDMA}-schemes for tree-routing in data intensive wireless sensor networks, in Proceedings of the IEEE International Conference on Mobile {Adhoc} and {Sensor} {Systems} (MASS), Pisa, Italy, Oct 2007.

\bibitem[43]{juc_energy_2016}
I Juc, O Alphand, R Guizzetti, M Favre, A Duda, {Energy Consumption and Performance of {IEEE} 802.15.4e {TSCH} and {DSME}}, in Proceedings of the {IEEE} {Wireless} {Communications} and {Networking} {Conference} (WCNC), Doha, Qatar, Apr 2016.

\bibitem[44]{di_marco_effects_2013}
P Di Marco, C Fischione, F Santucci, K Johansson, {Effects of {Rayleigh}-Lognormal Fading on {IEEE} 802.15.4 Networks}, in Proceedings of the {IEEE} {International} {Conference} on {Communications} ({ICC}), Budapest, Hungary, Jun 2013.

\bibitem[45]{petsc-efficient}
S Balay, W D Gropp, L C McInnes, B F Smith, Efficient Management of Parallelism in Object Oriented Numerical Software Libraries, in Modern Software Tools in Scientific Computing, ed. by E. Arge, A. M. Bruaset, and H. P. Langtangen{}, Birkh{\"{a}}user, Boston, MA, USA, pp.~163--202, 1997.

\bibitem[46]{petsc-user-ref}
S Balay, S Abhyankar, M F Adams, J Brown, P Brune, K Buschelman, L Dalcin, V Eijkhout, W D Gropp, D Kaushik, \textit{et al.}, {PETS}c Users Manual,  Technical Report ANL-95/11 - Revision 3.7,  (2016).

\bibitem[47]{chakravarti_characterization_1975}
I M Chakravarti, On a Characterization of Irreducibility of a Non-Negative Matrix. Linear Algebra and its Applications 10(2), 103--109 (1975).

\bibitem[48]{stewart_introduction_1994}
W J Stewart, Introduction to the {Numerical} {Solution} of {Markov} {Chains} (Princeton University Press, Princeton, NJ, USA, 1994).


\end{thebibliography}


\end{backmatter}
\end{document}